\documentclass[authoryear,final,3p]{elsarticle}

\usepackage{amssymb}
\usepackage{amsmath}

\usepackage{graphicx}%
\usepackage{multirow}%
\usepackage{amsmath,amssymb,amsfonts}%
\usepackage{amsthm}%
\usepackage{mathtools}
\usepackage{mathrsfs}%
\usepackage[title]{appendix}%
\usepackage{xcolor}%
\usepackage{textcomp}%
\usepackage{manyfoot}%
\usepackage{booktabs}%
\usepackage{algorithm}%
\usepackage{algorithmicx}%
\usepackage{algpseudocode}%
\usepackage{listings}%
\usepackage{commath}
\usepackage{hyperref}
\hypersetup{colorlinks  = true}
\usepackage{xurl}

\usepackage{cleveref}

\usepackage{changepage}

\usepackage{tikz}
\usetikzlibrary{bayesnet}

\journal{Epidemics}

\begin{document}

\begin{frontmatter}

\title{Bayesian inference for disease transmission models informed by viral dynamics}

\author{Dylan J. Morris\corref{cor1}\fnref{fn1}}
\ead{dylan.morris@adelaide.edu.au}

\author{Lauren Kennedy\fnref{fn1}}

\author{Andrew J. Black\fnref{fn1}}

\cortext[cor1]{Corresponding author}

\affiliation[fn1]{
  organization={School of Mathematical Sciences, Adelaide University},
  city={Adelaide},
    postcode={5005}, 
state={South Australia},
country={Australia}
}

\begin{abstract}
Infectious disease dynamics operate across multiple biological scales, with within-host viral dynamics being a key driver of between-host transmission. 
However, while models that explicitly link these scales exist, none have been developed with statistical inference as a primary goal.
In this paper we propose a multiscale model that jointly captures heterogeneous individual-level viral load trajectories and stochastic household transmission, and develop efficient inference methods to fit it to data. 
Since full joint inference is computationally difficult, we employ a cut approach that passes information from the within-host to the between-host model but not vice versa. 
This enables the data on viral loads to inform the transmission parameters such as the infection times and symptom onset thresholds. 
We evaluate the framework on simulated household outbreak data, assessing parameter recovery, computational efficiency, and the effect of viral load sampling frequency on inference quality. Parameter recovery is unbiased when the sampling frequency of the viral loads is high enough.
When sampling is sparse, some bias is introduced, but incorporating external viral load data can mitigate this.
\end{abstract}


\end{frontmatter}

\section{Introduction}\label{sec:intro}

Infectious disease dynamics occur across multiple scales 
\citep{almoceraMultiscaleModelWithinhost2018,gariraPrimerMultiscaleModelling2018,goyalMultiscaleModellingReveals2022,serenaReviewMultilevelModeling2023}. 
At the within-host scale, viral particles enter the body and infect cells, producing further virus \citep{perelsonModellingViralImmune2002,baccamKineticsInfluenzaVirus2006}. As populations of free virus and infected cells grow, driving up the viral load, the host becomes infectious. At the between-host scale, infected individuals disperse virus into the environment, infecting other individuals and the cycle continues \citep{almoceraMultiscaleModelWithinhost2018,gariraPrimerMultiscaleModelling2018}.
Standard practice is to model these two scales separately, exploiting the orders-of-magnitude difference in their timescales. 
However, as more data on within-host viral dynamics becomes available from polymerase chain reaction (PCR) testing, it can potentially be used to inform our understanding of disease dynamics at the population level, for instance, by helping to resolve individual infection times.
In addition, the biology relating viral load, immune response, and infectiousness is poorly understood \citep{handelCrossingScaleWithinhost2015,keVivoKineticsSARSCoV22021,marcQuantifyingRelationshipSARSCoV22021,puhachSARSCoV2ViralLoad2023} and multiscale models---if they can be fitted to data---can aid our understanding of these interactions \citep{gariraPrimerMultiscaleModelling2018,doranMathematicalMethodsScaling2023}. 

Fitting such models to data, however, is challenging. 
Calculating the likelihood requires integrating over latent viral trajectories for each individual, coupled through a non-linear transmission process and observed only indirectly through symptom onset and viral testing data.
This high-dimensional, partially observed, structure typically renders likelihood-based inference computationally intractable. As a result, existing multiscale approaches that link transmission to within-host viral dynamics primarily focus on forward simulation \citep{wangMultiscaleModelCOVID192022,tatsukawaAgentbasedNestedModel2023,yinAccurateStochasticSimulation2025}.
In this paper we take the next steps and propose a Bayesian framework for the inverse problem of parameter inference
for a combined within and between-host model.
We assume data arising from household transmission studies, which provide a natural setting for collecting high-resolution viral load and symptom onset data in a closed population \citep{blackCharacterisingPandemicSeverity2017, boddingtonEpidemiologicalClinicalCharacteristics2021,houseInferringRisksCoronavirus2022,marcatoLearningsAustralianFirst2022}.
We apply our method to simulated datasets comprising varying numbers of households, evaluating parameter recovery and computational efficiency. We also examine how the frequency of viral load sampling (daily vs every $r$ days) affects inference. 
Leveraging the hierarchical structure of the within-host model, we further show that incorporating external viral load data when fitting the model mitigates bias caused by sparse sampling.

To the best of our knowledge, this is the first study to propose a tractable inference method for jointly modelling these data sources. 
To achieve this we have adopted several model approximations and computational techniques.
To account for individual-level heterogeneity in viral dynamics and transmission, we adopt a hierarchical Bayesian framework, which allows information to be shared across individuals and households \citep{gelmanBayesianDataAnalysis}. 
This enables estimation of transmission parameters while preserving individual-level variation in infection timing and infectiousness.
Stochasticity in the within-host dynamics is included---without incurring the computational cost of full stochastic simulation---by employing a semi-stochastic hybrid approach \citep{morrisComputationRandomTimeshift2024,morrisRandomTimeshiftApproximation2025a}. This approximates sample paths of an underlying stochastic model by applying a random translation in time to deterministic solutions of the model. This translation is referred to as the random time-shift \citep{barbourEscapeBoundaryMarkov2015}, and the distribution can be estimated using methods from our previous work \citep{morrisComputationRandomTimeshift2024}. 
Modelling the viral load trajectories using this method captures key effects of early-phase noise on the measurable dynamics, particularly variability in peak timing, while incurring only a fraction of the expense of full simulation. 
It also makes aspects of the likelihood calculations more tractable \citep{morrisRandomTimeshiftApproximation2025a}. 

To overcome the complex posterior geometry that arises from full joint inference across both model scales, we adopt a cut-inference approach \citep{liuGeneralFrameworkCutting2025}. 
This is consistent with a nested multiscale modelling framework in which within-host viral dynamics inform between-host transmission but not vice versa \citep{gariraPrimerMultiscaleModelling2018}.
This simplifies inference and naturally accommodates the inclusion of external within-host data, allowing out-of-sample viral load observations to inform individual infectiousness profiles through the hierarchical model. 
To facilitate uncertainty propagation between the two stages, without carrying forward full posterior draws, we approximate the marginal posterior densities for within-host parameters with a Gaussian mixture model.

Household epidemic studies provide a natural setting for our framework \citep{blackCharacterisingPandemicSeverity2017,boddingtonEpidemiologicalClinicalCharacteristics2021,houseInferringRisksCoronavirus2022,marcatoLearningsAustralianFirst2022}. In particular, first-few-X (FFX) studies recruit the households of the first confirmed cases during an outbreak \citep{mcleanPandemicH1N120092010}, collecting data via PCR tests and symptom diaries that together provide some of the most complete epidemiological data available in early outbreak settings \citep{FirstFewCases2021,marcatoOngoingValueFirst2022}. 
Households form small, largely closed, populations yielding multiple independent realisations of transmission, and repeated PCR testing allows individual viral trajectories to be estimated.
With tractable inference for multiscale models, these features make it possible to reconstruct transmission chains and quantify the relationship between viral load and infectiousness.

All the code to replicate this study is available in the following GitHub repository:
\url{https://github.com/djmorris7/Bayesian-inference-for-disease-transmission-models-informed-by-viral-dynamics}.

\section{Methods}\label{sec:methods}

This section begins with specification of the within-host and between-host models before describing how we connect the two through an intensity function for an individual's infectiousness. 
We then present a forward simulation algorithm that generates data representative of a household study. Following this, we describe a Bayesian inference algorithm that performs the inverse operation---inferring the model parameters from data.

\subsection{Within-host model}\label{sec:within_host_model}

For our within-host model we adopt the target-cell-limited (TCL) with eclipse-phase model \citep{baccamKineticsInfluenzaVirus2006,smithInfluenzaVirusInfection2018,heTemporalDynamicsViral2020}. 
This describes the basic cell infection and viral production cycles within a host following infection with a typical infectious disease. 
More complex models exist that include immune responses \citep{challengerModellingUpperRespiratory2022,zitzmannHowRobustAre2024}, however, in this work we restrict ourselves to this simpler model, as our main contribution is to demonstrate how inference can be performed on a coupled within- and between-host model.
Replacing this with a more complex model is straightforward as long as it outputs the viral load through time.
Important theory and key results required to construct the within-host component of our model are included here, but we refer the reader to \citet{morrisRandomTimeshiftApproximation2025a} for further mathematical details. 

The TCL model tracks the numbers of susceptible target cells, $A(t)$, cells in the eclipse phase, $B(t)$, infected cells, $C(t)$, and free virus, $V(t)$, in a volume over which the within-host infection process occurs. 
Suppose the infection is initiated at some time $u$, then the state of the process at time $t \ge u$ is specified by the vector $(A(t), B(t), C(t), V(t))$. 
We model the progression of infection within a host using a hybrid model that combines deterministic and stochastic dynamics \citep{barbourEscapeBoundaryMarkov2015,morrisComputationRandomTimeshift2024,morrisRandomTimeshiftApproximation2025a}.
The deterministic component is given by the following system of ODEs \citep{baccamKineticsInfluenzaVirus2006,zitzmannHowRobustAre2024,morrisRandomTimeshiftApproximation2025a}
\begin{align}\label{eq:odes}
    \begin{split}
        \dod{\tilde A(t)}{t} &= -\beta \tilde A(t) \tilde V(t), \\ 
        \dod{\tilde B(t)}{t} &= \beta \tilde A(t) \tilde V(t) - k \tilde B(t), \\ 
        \dod{\tilde C(t)}{t} &= k \tilde B(t) - \delta \tilde C(t), \\ 
        \dod{\tilde V(t)}{t} &= \rho \tilde C(t) - c \tilde V(t),
    \end{split} 
\end{align}
where tildes denote these being deterministic solutions (i.e.~distinct from their stochastic counterparts defined in the next paragraph). 
The initial condition of the ODE is $(A_0 - 1, 1, 0, 0)$, where $A_0 = \tilde{A}(u) = 8\times 10^7$ is the initial number of susceptible cells, based on \citet{morrisRandomTimeshiftApproximation2025a}. 
In this work, interest is solely in the viral load through time, $V(t)$, and the parameters $\boldsymbol{\theta} = (R_0, k, \delta, \rho, c)$, where 
$$
R_0 = \frac{A_0 \beta \rho}{c \delta}
$$
is the within-host reproduction number \citep{zitzmannHowRobustAre2024}. 

The hybrid stochastic model is constructed from this deterministic solution as follows. Assuming fixed parameters $\boldsymbol{\theta}$ and infection time $u$, we obtain a solution $\tilde{V}(t)$ for the viral load through time as the solution to
Eq.~\eqref{eq:odes}. 
Our hybrid model is defined as
\begin{equation}\label{eq:hybrid_model}
    V(t) = \begin{cases}
        \tilde{V}(t + \tau) \quad \text{if } t + \tau > u \text{ and } t > u \\ 
        0 \quad \text{otherwise},
    \end{cases}
\end{equation}
where $\tau$ is a temporal translation to the deterministic solution and is referred to as the random time-shift \citep{barbourEscapeBoundaryMarkov2015,morrisComputationRandomTimeshift2024}. 
The random variable $\tau$ depends on the model parameters $\boldsymbol{\theta}$, but this is suppressed for clarity. 
Conditional on non-extinction, the probability density function (PDF) of $\tau$ can be accurately approximated by a member of a family of transformed generalised Gamma distributions \citep{morrisComputationRandomTimeshift2024}.
As in our previous work \citep{morrisRandomTimeshiftApproximation2025a}, we learn the mapping from $\boldsymbol{\theta}$ to timeshift distribution parameters (a scale and two shape parameters of the generalised Gamma) using a neural network, enabling us to amortise the cost of solving for the parameters \citep{marinoGeneralMethodAmortizing2018,amosTutorialAmortizedOptimization2023}.
The cases given in Eq.~\eqref{eq:hybrid_model} reflect the domain over which the ODEs are valid ($t \ge u$) and the interpretation of $t$ as calendar time \citep{morrisRandomTimeshiftApproximation2025a}.
We also define $\nu(t) = \log_{10} V(t)$ which is a more natural measurement scale for these models \citep{hayEstimatingEpidemiologicDynamics2021,challengerModellingUpperRespiratory2022,zitzmannHowRobustAre2024}.

Together, the ODE system Eq.~\eqref{eq:odes} and the neural network enabled approximation to the distribution of the random time-shift $\tau$ provide a method of sampling realisations of viral load trajectories for each individual. 
These hybrid trajectories closely approximate those of a fully stochastic within-host model, yet are computationally inexpensive to generate.
The simulated viral load trajectories feed directly into the between-host transmission model described in Section~\ref{sec:between_host_model}.
Example behaviour and trajectories of the hybrid within-host model are shown in Appendix Fig.~\ref{fig:example_trajectories}.
The approximation breaks down when populations are small; however, such small viral loads lie below the detection limit and this regime is therefore largely unobservable in practice \citep{kisslerstephenm.ViralDynamicsSARSCoV22021,challengerModellingUpperRespiratory2022,zitzmannHowRobustAre2024}.
Consequently, the approximation is well suited to the range of interest.  
The likelihood of trajectories under this approximation is also straightforward to evaluate, which facilitates further calculations \citep{morrisComputationRandomTimeshift2024}.

Measurement of the viral load, via PCR testing, is modelled as follows. 
Let $i$ label the individuals and recall that $t$ denotes calendar time (i.e.~time measured from some fixed reference point), with $u_i$ denoting the calendar time at which individual $i$ is infected).
Each individual is assumed to have $J_i$ measurements, where the number of measurements varies across individuals due to practical constraints such as staggered recruitment times, irregular testing schedules, or early cessation of sampling following recovery or loss to follow-up.
For the $i^{\text{th}}$ individual, measurements are taken at times $\boldsymbol{t}_i = (t_{i,1}, \ldots, t_{i,J_i})$, where we allow each individual to have measurements at different points in time. We denote the $j^{\text{th}}$ measurement for the $i^{\text{th}}$ individual as taken at time $t_{i,j}$.

While ideally we would observe the true log-10 viral load $\nu_i(t_{i,j})$, in practice we can only measure a noisy realisation of it, denoted $y_{i,j} := y_{t_{i,j}}$. 
Furthermore, the data are left-censored at a detection limit $\gamma$, meaning that if the noisy measurement falls below $\gamma$, we record only $\gamma$ rather than the true value (i.e. $y_{i,j} = \gamma$ means $y_{i,j} \le \gamma$). 
We assume $\gamma = 2.65$ based on previous work \citep{morrisRandomTimeshiftApproximation2025a}. 
The relationship between the true and observed viral load is therefore
\begin{equation*}
    y_{i,j} = \begin{cases}
        \nu_i(t_{i,j}) + \epsilon_{i,j} & \text{if }  \nu_i(t_{i,j}) + \epsilon_{i,j} > \gamma \\
        \gamma & \text{if } \nu_i(t_{i,j}) + \epsilon_{i,j} \leq \gamma,
    \end{cases}
\end{equation*}
where $\epsilon_{i,j} \sim \mathcal{N}(0, \kappa^2)$ is a measurement error term, and $\kappa$ is the standard deviation.

\subsection{Between-host transmission model}\label{sec:between_host_model}

Here we describe how the within-host model is coupled to a transmission model. Between-host transmission can be represented either by a compartmental model, which tracks the flow of individuals between discrete disease states such as susceptible, infected, and recovered \citep{allenPrimerStochasticEpidemic2017}, or by a generation time framework, which characterises transmission through the distribution of times between successive infections in a chain of transmission \citep{svenssonNoteGenerationTimes2007}. We adopt the latter because it couples naturally to the within-host model. Infectiousness is modelled as a function of viral load \citep{handelCrossingScaleWithinhost2015}, so the within-host trajectory directly defines an infectiousness profile, and the generation time distribution can be derived from it without collapsing the continuous dynamics into discrete disease states.

\begin{figure}[!ht]
    \centering
    \includegraphics{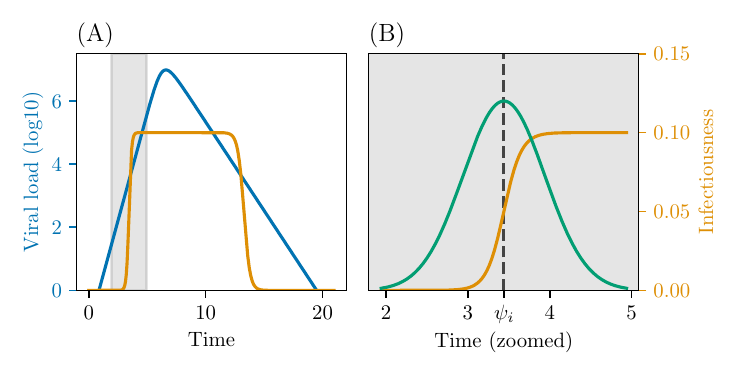}
    \caption{(A) A realisation of the within-host viral load trajectory (blue), with the corresponding force of infection (orange) overlaid. 
    (B) A zoomed-in view of the shaded region in panel (A), showing the force of infection (orange) and assumed symptom onset time distribution overlaid in green. The mean of the symptom onset time distribution, $\psi_i$, is indicated by the dashed black vertical line. 
    Note that axes are colour coded to their respective lines and that the overlaid density is shown on an arbitrary scale for visualisation purposes.}
    \label{fig:within_between_host_link}
\end{figure}

The force of infection is defined as the rate at which susceptible individuals become infected due to contact with infectious individuals \citep{keelingModelingInfectiousDiseases2008}.
Assuming we have individual $i$'s viral load, $V_i(t)$, we model the per-individual contribution to the force of infection as a Hill function of viral load.
Let the force of infection from a single infectious individual, acting on susceptible individuals, at time $t$, be given by
\begin{equation}\label{eq:lambda}
    \lambda_i(t) = \eta \left(\dfrac{V_i(t)^\alpha}{V_i(t)^\alpha + \Omega_i^\alpha} \right),
\end{equation}
where $\eta$ is the transmission rate, $\Omega_i$ is the half saturation (viral load) threshold, and $\alpha$ corresponds to the Hill coefficient \citep{martinez-corralHillFunctionUniversal2024}.
This Hill function maps $V_i(t)$ to a value in $[0, 1]$ \citep{martinez-corralHillFunctionUniversal2024}, so as $V_i(t)$ rises and subsequently falls over the course of infection, the force of infection increases and decreases accordingly.
For simplicity in our work, we fix $\alpha = 3$.
This produces a smooth increase from low infectiousness to peak infectiousness in the Hill function over the course of approximately a day for test parameter sets.
Note that this formulation produces infectiousness dynamics that are broadly consistent with standard compartmental models such as the susceptible-exposed-infectious-recovered (SEIR) model \citep{keelingModelingInfectiousDiseases2008}, but avoids the biologically unrealistic assumption of an abrupt transition from non-infectious to fully infectious.
Instead, infectiousness increases and decreases smoothly as a function of viral load.
An example of this function is shown alongside a typical viral load curve in Fig.~\ref{fig:within_between_host_link}A.
Note that because $V_i(t)$ is a stochastic process, the force of infection, $\lambda_i(t)$ is also stochastic. This means that there is variability in the timing and magnitude of the infectiousness across individuals. 

It is often preferable to work in log-10 scale, as this is the natural unit for viral loads. 
We therefore reparameterise the half-saturation threshold as $\omega_i = \log_{10} \Omega_i$. 
This has the additional practical benefit of placing $\omega_i$ on a numerically convenient scale for Markov chain Monte Carlo (MCMC) sampling, where its magnitude is comparable to that of other model parameters.

The final part of the model relates symptom onset to the viral load. 
We first define the time where the log-10 half saturation threshold is crossed for individual $i$ as 
$$
\psi_i = \arg\min_{t} \left\{ \nu_i(t) \ge \omega_i \right\}.
$$
We make the simplifying assumption that symptoms always occur, i.e., that $V_i(t)$ always crosses $\omega_i$, so that $\psi_i$ is well-defined for all individuals.
To accommodate both heterogeneity arising from the process, and the data collection itself (symptoms manifest in a range of ways and onset is not a binary variable), we model the reported symptom onset time as a noisy realisation,
\begin{equation}
    d_i \sim \mathcal{N}(\psi_i, 0.5^2), 
\end{equation} 
where we have taken the standard deviation to be $0.5$, so that the 95\% confidence interval reflects an approximate $\pm 1$ day window of uncertainty about the crossing time.
This formulation does not require the reported symptom onset time to be after the infection time, as reporting errors and/or inconsistencies in symptom diaries could cause such a result. 
Fig.~\ref{fig:within_between_host_link}B shows how the symptom onset time is linked to the viral load curve for a single individual.

\subsection{Data generation}\label{sec:data}

To ground this framework in a concrete setting, we consider household transmission studies as our primary application.
In a household study, each household is treated as an independent population where contact between household members is assumed to be frequent and well characterised,
while between-household transmission is assumed negligible over the observation window.
We therefore model each household outbreak as an independent realisation of the same underlying transmission process.

Suppose the study comprises \(H\) households, and let \(\mathcal{H} = \{1,\ldots,H\}\) denote the set of household identifiers.
Consider the $h^{\text{th}}$ household of $N_h$ individuals, and suppose that $M_h \le N_h$ are infected over the course of the outbreak. 
We further assume that the outbreak is complete, so that $M_h$ is the final number of infected individuals. 
Finally, we assume a closed household and that the outbreak is initiated by a single infection. 

We extend the model notation from Sections~\ref{sec:within_host_model} and \ref{sec:between_host_model}, adding a household index $h$, so that $y_{h,i,j}$ denotes the $j$-th viral load observation for individual $i$ in household $h$, and $d_{h, i}$ denotes their symptom onset time.
Model parameters follow a similar convention. 
Within-host viral dynamics are governed by the within-host model described in Section~\ref{sec:within_host_model}, with individual-specific parameters, 
$$
\boldsymbol{\theta}_{h,i} = (R_{0_{h,i}}, \delta_{h,i}, \rho_{h,i}),
$$ 
controlling their viral load trajectory.
For identifiability, we fix some parameters of the within-host model to be constant across individuals. 
Specifically, we set the eclipse phase parameter \(k_{h,i} = 4\) and viral clearance rate \(c_{h,i} = 10\) for all \(h\) and \(i\), following \citet{zitzmannHowRobustAre2024}; these are therefore excluded from $\boldsymbol{\theta}_{h, i}$.
We stratify by household in this way because, as we will show in Section~\ref{sec:between-host_inference}, households form a natural partition of the likelihood function for the between-host model. 

We have two sources of observed data in our modelling process. 
The first are the log-10 viral loads for each infected individual at each measurement point.
We collect the viral load data for household $h$ in $\mathcal{Y}_h = \{\boldsymbol{y}_{h,1}, \boldsymbol{y}_{h,2}, \ldots, \boldsymbol{y}_{h,M_h} \}$, where $\boldsymbol{y}_{h,i} = (y_{h,i,1}, \ldots, y_{h,i,J_{h,i}})$ denotes the vector of viral load observations for individual $i$ in household $h$, and $J_{h,i}$ is the number of observations for that individual. 
The second is the symptom onset times for each individual. 
Let $\boldsymbol{d}_h = (d_{h, 1}, \ldots, d_{h, M_h})$ denote the reported symptom onset times of the infected individuals in household $h$, as described in Section~\ref{sec:between_host_model}.
Then the combined observed information for household $h$ comprises repeated testing measurements for each individual and symptom onset times for infected individuals, we then denote the household set of data as $\{ \boldsymbol{d}_h, \mathcal{Y}_h \}$.

In our model, individual viral loads progress over time, so transmission rates are time-varying and the model is therefore non-Markovian. 
To simulate from such a the model we must track the full history of each individual's infection.
To do this we split the simulator into two components: an individual-level viral load simulator and a population-level transmission simulator.
We describe each briefly below, leaving complete algorithmic details to Appendix~\ref{sec:simulation}.
In the individual level simulator, we generate viral load trajectories for an individual and the reported symptom onset time. 
From the viral load trajectory we compute the times of infection attempts that this individual makes, where an infection attempt represents a potential transmission event whose rate is governed by the force of infection $\lambda_{h, i}(t)$. 

The population simulator takes this information and uses rejection sampling to determine which attempts result in actual infections, accepting each with probability $S(t) / (N_h - 1)$ where $S(t)$ is the number of susceptibles at time $t$. 
This reflects frequency-dependent transmission whereby the per-contact infection rate depends on infectiousness, but actual infections also require a susceptible contact to be made.
Stochasticity in the transmission process arises from both variability in individual viral load trajectories and the Poisson process governing transmission events.
The output of the simulation is a dataset of viral load trajectories and reported symptom onset times. 
As the observed viral loads do not influence transmission dynamics, they can be simulated independently from the viral load trajectory in a post-processing step.

\begin{figure}[!ht]
    \centering
    \includegraphics{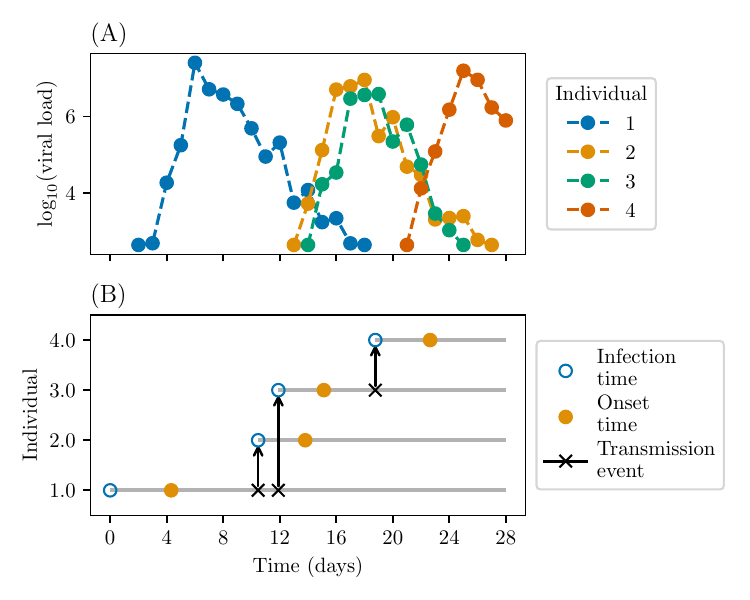}
    \caption{
        An example outbreak in a household of size 4 over $28$ days. 
        (A) The log-10 viral load time series for each member of the household, coloured by household member. 
        We show dashed lines linking the points for ease of view but the data is only the discretely observed solid coloured dots.
        (B) Household level information with grey lines indicate when an individual is infected. 
        Open blue dots denote individual infection times, and solid red dots denote reported symptom onset times. 
        Note that the infection times are unobserved. 
        Black crosses indicate transmission events, and vertical arrows show the (unknown) infector–infectee pairs.
    }
    \label{fig:household_line_list}
\end{figure}

Fig.~\ref{fig:household_line_list} illustrates a single household outbreak as an example. 
Under the simplifying assumption of daily PCR testing, we observe viral load time series for each individual which are shown in Fig.~\ref{fig:household_line_list}A.
The solid dots mark observed data, where we assume daily collection of data.
This is an idealised data collection scenario. 
In our results we investigate more realistic scenarios where the sampling of the viral load trajectories is sparser.
Fig.~\ref{fig:household_line_list}B shows the transmission chain within the household. 
The observed data consist of the reported symptom onset times for each individual (solid orange dots). 
Neither infection times nor infection chains are directly observed.

\subsection{Inference methodology}\label{sec:inference}

Section~\ref{sec:overview} outlines the overall inference strategy and its motivation.
Section~\ref{sec:notation} introduces the notation used throughout this section and presents a plate diagram illustrating the model structure.
Subsequent sections detail the inference procedure and the derivation of the likelihood components.

\subsubsection{Overview of inference method}\label{sec:overview}

The combined within- and between-host model defines a joint posterior distribution over all model parameters. 
In principle, inference could be performed on the full joint posterior. 
However, this is challenging in practice for two reasons. 
First, some model components are not amenable to automatic differentiation and therefore require custom MCMC methods. 
Second, the full joint posterior would couple transmission events (through the between-host model) to viral trajectories (the within-host model). This would mean that transmission event data and the priors in the between-host model could inform associated within-host parameters. If both models were perfectly specified, then this would allow allow symptom onset and transmission data to inform within-host kinetics. However, jumping to this approach directly substantially complicates model interpretation and validation, as well as model computation. 

Given the computational and inferential challenges associated with the full joint posterior, we instead adopt a modular approach motivated by the unidirectional coupling based nested multiscale modelling (UNID-NSM) framework \citep{gariraPrimerMultiscaleModelling2018}. 
In this framework, information flows strictly from the within-host model to the between-host model, consistent with the forward simulation where within-host dynamics drive transmission. 
Formally, this corresponds to limiting information from the between-host component to the within-host parameters \citep{liuGeneralFrameworkCutting2025}.
To achieve this, inference is performed in two stages. 
First, inference is conducted using the within-host model and viral load measurements. 
Posterior samples of parameters that appear in the between-host model (such as $\boldsymbol{\theta}_{h,i}$) are then approximated using Gaussian mixture distributions. These Gaussian mixture distributions are then used as priors for the between-host model. This process allows information to flow from the within-host model to the between-host model but doesn't allow information flow in the opposite direction. Even though information flows between the two model components, the between-host model still updates the priors derived from the within-host model, limiting the impact of any model mis-specification in the within-host model. 

\subsubsection{Setup and notation}\label{sec:notation}

We collect within-host parameters for individuals in household \(h\) in the set
\[
\Theta_{h} = \{ \boldsymbol{\theta}_{h,1}, \ldots, \boldsymbol{\theta}_{h,M_h} \}.
\]
where $\boldsymbol{\theta}_{h,i} = (R_{0_{h, i}}, \delta_{h, i}, \rho_{h, i})$.
Within-host parameters are drawn from distributions governed by the hyper parameters
\[
\boldsymbol{\phi}
=
(\mu_{R_0}, \sigma_{R_0},
 \mu_{\delta}, \sigma_{\delta},
 \mu_{\rho}, \sigma_{\rho},
 \kappa),
\]
for example \(R_{0_{h,i}} \sim \mathcal{N}_{+}(\mu_{R_0}, \sigma_{R_0})\).
These hierarchical relationships are described in full in Section~\ref{sec:within-host_inference}.

Let \(\boldsymbol{u}_{h}\) denote the (not necessarily ordered) vector of infection times for individuals in household \(h\).  
Let \(\boldsymbol{\omega}_{h}\) denote the vector of detection thresholds for individuals in household \(h\), which are assumed to depend on hyperparameters \(\mu_{\omega}\) and \(\sigma_{\omega}\).
Finally, let
\[
\boldsymbol{\zeta} = (\eta, \mu_{\omega}, \sigma_{\omega}),
\]
denote the between-host transmission parameters. 
Priors for the between-host parameters are detailed in Section~\ref{sec:between-host_inference}.
Some parameters, such as $\boldsymbol{\theta}_{h, i}$ and $u_{h, i}$, appear in both stages of the model. 
To distinguish between them, we use superscripts $(W)$ and $(B)$ to denote parameters inferred in the within-host and between-host models, respectively. 
For example, $\boldsymbol{\theta}_{h,i}^{(W)}$ denotes the within-host parameter vector for individual $i$ in household $h$.

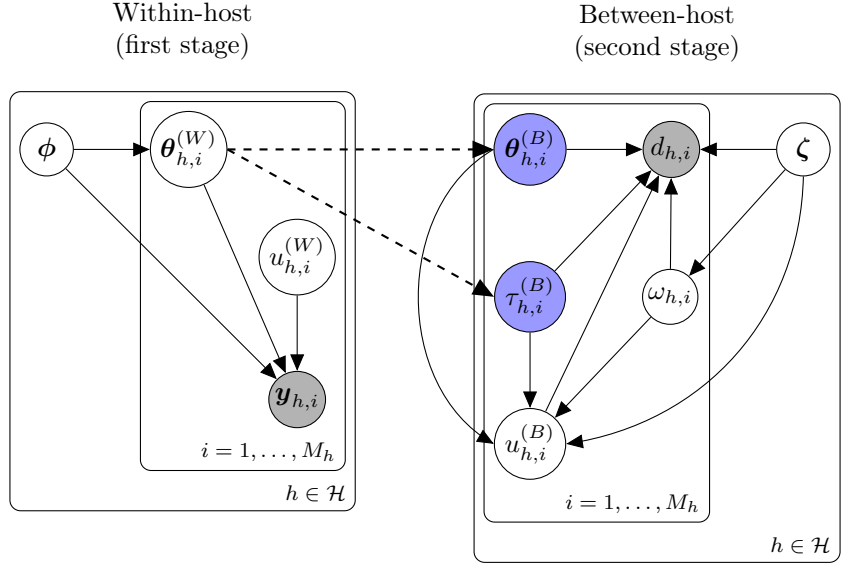
\begin{figure}[!ht]
    \centering
    \begin{tikzpicture}
        \node[latent]                   (phi)       {$\boldsymbol{\phi}$};
        \node[latent, right=of phi]     (theta)     {$\boldsymbol{\theta}_{h,i}^{(W)}$};
        \node[latent, below right=of theta]   (iota)      {$u_{h,i}^{(W)}$};
        \node[obs, below=of iota, fill=gray!60]       (Y)         {$\boldsymbol{y}_{h,i}$};
    
        \edge {phi} {theta};
        \edge {theta, iota, phi} {Y};
    
        \plate {indiv1} {(theta)(iota)(Y)} {$i=1,\dots,M_h$};
        \plate {house1} {(indiv1)(phi)} {$h \in \mathcal{H}$};
        \node[align=center, above=0.3cm of house1.north] (wh_label) {Within-host \\ (first stage)};
    
        \node[latent, right=of theta, xshift=2.5cm, fill=blue!40] (thetaS) {$\boldsymbol{\theta}_{h,i}^{(B)}$};
        \node[latent, below=of thetaS, fill=blue!40]  (tauS)  {$\tau_{h,i}^{(B)}$};
        \node[latent, below=of tauS]      (iotaS) {$u_{h, i}^{(B)}$};
        \node[latent, right=of tauS]  (omega)    {$\omega_{h, i}$};
        \node[obs, right=of thetaS, fill=gray!60]       (d)     {$d_{h,i}$};
        \node[latent, right=of d]       (zeta)  {$\boldsymbol{\zeta}$};
    
        \edge {thetaS, tauS, omega, iotaS} {d};
        \edge {zeta} {omega};
        \edge {zeta} {d};
        \edge {omega} {iotaS};
        \edge {tauS} {iotaS};

        \draw[->] (thetaS.west) to[bend right=60] (iotaS.west);
        \draw[->] (zeta.south)  to[bend left=40]  (iotaS.east);
    
        \plate {indiv2} {(thetaS)(tauS)(omega)(d)(iotaS)} {$i=1,\dots,M_h$};
        \plate {house2} {(indiv2)(zeta)} {$h \in \mathcal{H}$};
        \node[align=center, above=0.3cm of house2.north] (bh_label) {Between-host \\ (second stage)};
    
        \draw[->, dashed, thick] (theta.east) -- (thetaS.west);
        \draw[->, dashed, thick] (theta.east) -- (tauS.west);
    \end{tikzpicture}
    \caption{
        Plate diagram illustrating the two-stage within-host and between-host inference approach.
        Solid arrows denote conditional dependence in the generative model.
        Dashed arrows indicate dependencies between the first-stage (within-host) and second-stage (between-host) parameters, arising from the approximate prior constructed from the first-stage posterior (see Section~\ref{sec:surrogate_construction}). 
        Blue nodes in the between-host plate represent quantities informed by this approximate prior. 
        The data are shown as grey shaded nodes.
        Inner plates correspond to individual-level quantities within a household, while outer plates correspond to household-level quantities and shared hyperparameters.
    }
    \label{fig:plate_diagram}
\end{figure}

The general relationships between parameters, data, and the flow of information in the model are shown in the plate diagram in Fig.~\ref{fig:plate_diagram}.
Explicit details of these relationships including distributions, parameters, and their flow into the likelihood functions, are covered in Sections~\ref{sec:within-host_inference} (within-host inference) and \ref{sec:between-host_inference} (between-host inference).
The dashed lines indicate the flow of information between the two modules which is described in Section~\ref{sec:surrogate_construction}.

\subsubsection{First stage inference: within-host scale}\label{sec:within-host_inference}

The first stage of the model considers only the log-10 viral load data and the within-host dynamics, ignoring transmission dynamics inside the household. 
We conduct inference on the within-host parameters by targeting the posterior
\begin{equation}\label{eq:within-host_posterior}
\begin{aligned}
    &p\left( 
        \left\{ \Theta_{h}^{(W)}, \boldsymbol{u}_{h}^{(W)} \right\}_{h \in \mathcal{H}},
        \boldsymbol{\phi}
        \,\middle|\,
        \left\{ \mathcal{Y}_{h} \right\}_{h \in \mathcal{H}}
    \right) \\ 
    &\propto 
    p(\boldsymbol{\phi}) \prod_{h \in \mathcal{H}} 
    \prod_{i = 1}^{M_h}
    p\left(
        \boldsymbol{y}_{h, i}
        \,\middle|\,
        \boldsymbol{\theta}_{h, i}^{(W)}, u_{h, i}^{(W)},  
        \boldsymbol{\phi}
    \right) 
    p \left( 
        \boldsymbol{\theta}_{h, i}^{(W)} \,\middle|\, \boldsymbol{\phi}
    \right) 
    p\left( u_{h, i}^{(W)}\right).
\end{aligned}
\end{equation}

Inference is performed using a Metropolis-within-Gibbs MCMC scheme, and the within-host modelling, preprocessing, and prior specification follow \citet{morrisRandomTimeshiftApproximation2025a} unless otherwise stated.
Viral load time series are shifted such that the timing of peak viral load occurs at $t = 0$. 
We truncate each time series to retain only observations above the limit of detection ($2.65$ log-10 viral load), along with any below-threshold observations necessary to define the endpoints. 
Time series were further truncated after three consecutive observations below this threshold. 
Under this convention, infection times $u_{h, i}^{(W)}$ are expressed relative to the peak viral load.

We assume that $R_{0_{h, i}}^{(W)}, \delta_{h, i}^{(W)}$ and $\rho_{h, i}^{(W)}$ for all $i$ and $h$, each follow truncated normal distributions with individual-specific hyperparameters. 
For example, $R_{0_{h, i}}^{(W)}$ is assigned the prior, 
$$
R_{0_{h, i}}^{(W)} \mid \mu_{R_0}, \sigma_{R_0} \sim \mathcal{N}_{+}(\mu_{R_0}, \sigma_{R_0}^2).
$$ 
We assume the following independent priors on the hyperparameters
\begin{align*}
    \mu_{R_0} &\sim \textrm{Gamma}\left(\dfrac{10}{3},\, 3\right), \\
    \mu_{\delta} &\sim \textrm{Gamma}\left(\dfrac{1.3}{0.05},\, 0.05\right), \\
    \mu_{\rho} &\sim \textrm{Gamma}\left(\dfrac{3}{0.3},\, 0.3\right), \\
    \sigma_{R_0} &\sim \mathcal{N}_{+}(0, 3^2), \\
    \sigma_{\delta} &\sim \mathcal{N}_{+}(0, 1^2), \\
    \sigma_{\rho} &\sim \mathcal{N}_{+}(0, 3^2).
\end{align*}
For the infection times we use a weakly informative prior
\begin{equation*}
    u_{h, i}^{(W)} \sim \textrm{Gumbel}(-7, 3).
\end{equation*}
In this stage of the model, the infection times are treated as auxiliary variables, i.e. variables introduced to facilitate likelihood evaluation and posterior sampling, rather than quantities of direct inferential interest. In the between-host part, these are re-inferred.

\subsubsection{Second stage inference: between-host model}\label{sec:between-host_inference}

The second stage of the model considers the symptom onset time, within-host dynamics and the transmission dynamics. 
We conduct inference over the parameters in this model by targeting the posterior 
\begin{equation}\label{eq:between_host_target}
\begin{aligned}
    &p\left( 
        \left\{ \Theta_{h}^{(B)}, \boldsymbol{u}_{h}^{(B)}, 
        \boldsymbol{\tau}_{h}^{(B)}, \boldsymbol{\omega}_{h} \right\}_{h \in \mathcal{H}},
        \boldsymbol{\zeta}
        \,\middle|\,
        \left\{ \boldsymbol{d}_{h} \right\}_{h \in \mathcal{H}}
    \right) \\
    &\propto \prod_{h \in \mathcal{H}} 
    \Bigg[p\left( 
            \boldsymbol{d}_{h} 
            \,\middle|\,
            \Theta_{h}^{(B)}, \boldsymbol{u}_{h}^{(B)}, 
            \boldsymbol{\tau}_{h}^{(B)}, \boldsymbol{\omega}_{h},
            \boldsymbol{\zeta}
        \right) \\
    &\quad \times p\left(
            \boldsymbol{u}_{h}^{(B)}
            \,\middle|\,
            \Theta_{h}^{(B)}, \boldsymbol{\tau}_{h}^{(B)},
            \boldsymbol{\omega}_{h}, \boldsymbol{\zeta}
        \right) \\
    &\quad \times \hat{p}\!\left(
            \Theta_{h}^{(B)}, \boldsymbol{\tau}_{h}^{(B)}
        \right) p\left( 
            \boldsymbol{\omega}_{h} \mid \boldsymbol{\zeta} 
        \right) p(\boldsymbol{\zeta}) \Bigg].
\end{aligned}
\end{equation}

Several parameters in the between-host model, namely $\Theta_{h}^{(B)}$, $\boldsymbol{u}_{h}^{(B)}$, and $\boldsymbol{\tau}_{h}^{(B)}$, relate to quantities in the within-host model (see Fig.~\ref{fig:plate_diagram}). 
The prior $\hat{p}\left(\Theta_h^{(B)}, \boldsymbol{\tau}_h^{(B)}\right)$ is an approximation of the posterior from the within host models, $\hat{p}\left(\Theta_h^{(W)}, \boldsymbol{\tau}_h^{(W)}\right)$, and hence allows the flow of information between the two models (described in detail in Section~\ref{sec:surrogate_construction}). 
As mentioned in Section~\ref{sec:within-host_inference}, the infection times 
$\boldsymbol{u}_h^{(W)}$ are treated as auxiliary parameters. 
In the between-host stage, they are re-inferred using symptom onset data and the transmission model, so that $\boldsymbol{u}_h^{(B)}$ is inferred given $\Theta_h^{(B)}$ and $\boldsymbol{\tau}_h^{(B)}$, independently of the corresponding within-host quantity $\boldsymbol{u}_h^{(W)}$.

Eq.~\eqref{eq:between_host_target} can be evaluated by factorising the contributions inside the product. 
The first term is the likelihood of observing the reported symptom onset times given the parameters. 
This term factorises over individuals
\begin{equation*}
    p\left(
        \boldsymbol{d}_h
        \,\middle|\,
        \Theta_h^{(B)}, \boldsymbol{u}_h^{(B)}, 
        \boldsymbol{\tau}_h^{(B)}, \boldsymbol{\omega}_{h},
        \boldsymbol{\zeta}
    \right) = \prod_{i = 1}^{M_h} p\left(
        d_i
        \,\middle|\,
        \boldsymbol{\theta}_{h, i}^{(B)}, u_{h, i}^{(B)}, 
        \tau_{h, i}^{(B)}, \omega_{h, i},
        \boldsymbol{\zeta}
    \right).
\end{equation*}
Conditional on the within-host parameters $\boldsymbol{\theta}_{h, i}^{(B)}$ and fixed $\tau_{h, i}^{(B)}$, the viral dynamics models $\nu_{h, i}(t)$ are deterministic. 
Given the fixed deterministic curve and fixed threshold $\omega_{h, i}$, we solve for the crossing time, $\psi_{h, i}$, as described in Section~\ref{sec:data}. 
Then the likelihood is simply
\begin{equation*}
    p\left(
        \boldsymbol{d}_h
        \,\middle|\,
        \Theta_h^{(B)}, \boldsymbol{u}_h^{(B)}, 
        \boldsymbol{\tau}_h^{(B)}, \boldsymbol{\omega}_{h},
        \boldsymbol{\zeta}
    \right) = \prod_{i = 1}^{M_h} \mathcal{N}_{\textrm{PDF}}(d_{h, i} \mid \psi_{h, i}, 0.5^2)
\end{equation*}
where $\mathcal{N}_{\textrm{PDF}}(d \mid \psi, 0.5^2)$ is the density function for a $\mathcal{N}(\psi, 0.5^2)$ random variable evaluated at $d$.

The next term in the product of Eq.~\eqref{eq:between_host_target},
$
p \left( \boldsymbol{u}_{h}^{(B)} \,\middle|\, \Theta_{h}^{(B)}, \boldsymbol{\tau}_{h}^{(B)}, \boldsymbol{\omega}_{h}, \boldsymbol{\zeta} \right),
$
is the transmission model that links the within-host and between-host dynamics.  
This can be thought of as an informative prior over the infection times conditional on individual viral profiles and the between-host dynamics. 
In particular it gives the probability of observing a sequence of infection times arising from a generation time model. 
Let $\boldsymbol{u}_h^\prime$ denote the sorted infection times for household $h$, i.e. $u^\prime_{h, i}$ refers to the $i$th infection event. 
Then, assuming the outbreak is observed over some time-horizon $T$,
\begin{equation}\label{eq:transmission_pdf}
    p \left( \boldsymbol{u}_h \,\middle|\, \Theta_h^{(B)}, \boldsymbol{\tau}_h^{(B)}, \boldsymbol{\omega}_{h}, \boldsymbol{\zeta} \right) = \mathcal{C}_h \prod_{i = 2}^{M_h} p \left( u^{\prime}_{h, i} \,\middle|\, \boldsymbol{u}^{\prime}_{h, 1:i-1}, \Theta_h^{(B)}, \boldsymbol{\tau}_h^{(B)}, \boldsymbol{\omega}_{h}, \boldsymbol{\zeta} \right)
\end{equation}
where 
$$
\mathcal{C}_h = \textrm{Pr}\left( \left\{ S(T) - S(u^{\prime}_{h, M_h}) = 0 \right\} \,\middle|\, \boldsymbol{u}^{\prime}_{h, 1:i-1}, \Theta_h^{(B)}, \boldsymbol{\tau}_h^{(B)}, \boldsymbol{\omega}_{h}, \boldsymbol{\zeta}\right)
$$
is the probability that there is no change in the susceptible count after the final infection and until the study duration $T$. 
The product term begins at $i = 2$ as $u^{\prime}_{h, i}$ denotes the infection time of the index case.  

The conditional terms in Eq.~\eqref{eq:transmission_pdf} are due to the sequential construction of infection times in a continuous-time transmission process. 
At any time $t$, each infected individual $i$ contributes a force of infection $\lambda_{h, i}(t)$ for transmitting infection to a susceptible contact, and the total force of infection in household $h$ is therefore  
\begin{equation*}
    \lambda_{h}^{\star}(t) = \sum_{i \in \mathcal{A}_{h, t}} \lambda_{h, i}(t),
\end{equation*}
where $\mathcal{A}_{h, t} = \{j : u_{h, j} < t\}$ is the set of infectious individuals at time $t$. 
The corresponding cumulative force of infectiousness is  
\begin{equation*}
    \Lambda_h^{\star}(t) = \int_{u^{\prime}_{h, 1}}^t \lambda_h^{\star}(s) \, \dif s
    = \sum_{i \in \mathcal{A}_{h, t}} \Lambda_{h, i}(t),
\end{equation*}
where $\Lambda_{h, i}(t)$ is the cumulative force of infection from individual $i$.  
In this process, the density for the $i$-th infection time is the product of the instantaneous rate that an infection occurs at $u^{\prime}_{i}$, and the probability of no infections between $u^{\prime}_{h, i-1}$ and $u^{\prime}_{h, i}$.
This yields the expression
\begin{align*}
    &p \left( u^{\prime}_{h, i} \,\middle|\, \boldsymbol{u}^{\prime}_{h, 1:i-1}, \Theta_h^{(B)}, \boldsymbol{\tau}_h^{(B)}, \boldsymbol{\omega}_{h}, \boldsymbol{\zeta} \right)    \\
    &= \frac{S(u^{\prime}_{h, i-1})}{N_h - 1} \lambda_h^{\star}(u^{\prime}_{h, i}) \exp{\left[- \frac{S(u^{\prime}_{h, i-1})}{N_h - 1} \left( \Lambda_h^{\star}(u^{\prime}_{h, i}) - \Lambda_h^{\star}(u^{\prime}_{h, i-1}) \right) \right]}, 
\end{align*}
where the factor $S(u^{\prime}_{h, i-1})/(N_h - 1)$ accounts for the probability that the next infection event must be one of the remaining $S(u^{\prime}_{h, i-1})$ susceptible individuals.
Similarly, the factor $\mathcal{C}_h$ is the probability that no further infections occur after the last observed infection in the household and the end of the observation period. 
If $u^{\prime}_{h, M_h}$ is the last infection time in household $h$, then
\begin{equation*}
    \mathcal{C}_h = \exp{\left[- \frac{S(u^{\prime}_{h, M_h})}{N_h - 1} \left( \Lambda_h^{\star}(T) - \Lambda_h^{\star}(u^{\prime}_{h, M_h}) \right) \right]}
\end{equation*}
is the probability that all remaining susceptible individuals avoid infection after $u^{\prime}_{h, M_h}$, accounting for the case where no further infections occur within the observation window.

The terms $p(\boldsymbol{\omega}_h \mid \boldsymbol{\zeta}) p(\boldsymbol{\zeta})$
are the priors over the between-host viral thresholds and the shared parameters for the between-host scale of the model.
The priors on between-host shared parameters are chosen to be weakly informative
\begin{align*}
    \eta &\sim \textrm{Exp}(1), \\ 
    \mu_{\omega} &\sim \textrm{Gamma}(8, 0.5), \\ 
    \sigma_{\omega} &\sim \mathcal{N}_{+}(0, 0.3^2). 
\end{align*}
We assume that the individual log-10 half saturations, $\omega_{h, i}$, are normally distributed (truncated to be non-negative), 
$$
\omega_{h, i} \mid \mu_{\omega}, \sigma_{\omega} \sim \mathcal{N}_{+}(\mu_{\omega}, \sigma_{\omega}^2).
$$
The hyper-priors for $\mu_{\omega}$ and $\sigma_{\omega}$ were chosen by sampling from the prior predictive distribution of the $\omega_{h, i}$ and asserting a reasonable range over the expected thresholds. 
The prior predictive distribution resulting from these choices has a 95\% confidence interval of $(1.64, 7.09)$ and median of $3.83$.
This covers a wide range of log-10 viral loads and is consistent with observed ranges of viral loads in published analyses \citep{baccamKineticsInfluenzaVirus2006,kisslerViralDynamicsAcute2021,challengerModellingUpperRespiratory2022,zitzmannHowRobustAre2024}.

\subsubsection{Approximate prior distribution construction}\label{sec:surrogate_construction}

We now describe the construction of $\hat{p}(\Theta_h^{(B)}, \boldsymbol{\tau}_h^{(B)})$, the approximate prior for the within-host parameters and time-shifts in the between-host model using the stage 1 (within-host) posterior samples. 
In the following we use hats to denote approximate densities.
We factorise the prior over individuals as
\begin{equation}\label{eq:w2b_prior}
    \hat{p} \left( \Theta_{h}^{(B)}, \boldsymbol{\tau}_{h}^{(B)}\right) = \prod_{i = 1}^{M_h} \hat{p} \left( \boldsymbol{\theta}_{h, i}^{(B)}, \tau_{h, i}^{(B)}\right),
\end{equation}
where each factor is defined by assigning the within-host posterior as the between-host prior.
Specifically, we first approximate the within-host posterior for individual $i$ as
\begin{equation}\label{eq:within_host_posterior_gmm_target}
    \hat{p}\left( \boldsymbol{\theta}_{h, i}^{(W)}, \tau_{h, i}^{(W)}\right) \approx p \left( \boldsymbol{\theta}_{h, i}^{(W)}, \tau_{h, i}^{(W)} \mid \{\mathcal{Y}_h\}_{h \in \mathcal{H}} \right),
\end{equation}
and then assign this as the prior for the corresponding between-host parameters via the relabelling
\[
\left( \boldsymbol{\theta}_{h, i}^{(B)}, \tau_{h, i}^{(B)} \right) \sim \hat{p}\left(\boldsymbol{\theta}_{h, i}^{(W)}, \tau_{h, i}^{(W)}\right).
\]
Eq.~\eqref{eq:w2b_prior} then serves directly as the prior in the between-host model.

Sampling the within-host model yields a collection of $n_s$ posterior draws
\begin{equation*}
    \left\{ \Theta_{h}^{(W, n)}, \boldsymbol{\phi}^{(n)} \right\}_{h \in \mathcal{H}}, 
    \qquad n = 1, \dots, n_s,
\end{equation*}
where $\Theta_{h}^{(W, n)}$ is the $n^{\text{th}}$ within-host draw for household $h$ and $\boldsymbol{\phi}^{(n)}$ is the corresponding vector of hyperparameters.
Samples obtained via the method described in Section~\ref{sec:within-host_inference} cannot be used directly for approximating the priors, as the force of infections $\lambda_{h,i}(t)$ require the time-shifts, which were marginalised out in the first stage. 
The induced marginal distribution of the time-shift for individual $i$ is obtained by integrating the conditional time-shift distribution over the within-host posterior of the parameters,
\begin{align*}
    &p\left(\tau_{h, i}^{(W)} \mid \{\mathcal{Y}_h \}_{h \in \mathcal{H}} \right) \\
    &= \int p\left(\tau_{h, i}^{(W)} \mid \boldsymbol{\theta}_{h, i}^{(W)}, \boldsymbol{\phi}\right) p\left(\boldsymbol{\theta}_{h, i}^{(W)}, \boldsymbol{\phi} \mid \{\mathcal{Y}_h \}_{h \in \mathcal{H}} \right) \dif \boldsymbol{\theta}_{h, i}^{(W)} \dif \boldsymbol{\phi}.
\end{align*}
Monte-Carlo samples of the time-shifts can be drawn from this density using random time-shift methods \citep{morrisComputationRandomTimeshift2024}, by first making a posterior draw $(\boldsymbol{\theta}_{h, i}^{(W, n)}, \boldsymbol{\phi}^{(n)})$, then constructing the time-shift distribution given $\boldsymbol{\theta}_{h, i}^{(W, n)}$ using the neural network approximation, and sampling $\tau_{h, i}^{(W, n)}$ for $n = 1, \dots, n_s$.

We fit the approximation Eq.~\eqref{eq:within_host_posterior_gmm_target} using a finite Gaussian mixture model (GMM) \citep{reynoldsGaussianMixtureModels2009}.
We choose this approach because it provides a flexible yet tractable way to summarise the posterior distributions. \citet{morrisRandomTimeshiftApproximation2025a} show that these posteriors, while they are unimodal, can have a slight skew. 
A GMM can capture this skew and the correlations between parameters in the posterior, which would be otherwise difficult to represent using a simple parametric distribution. 
At the same time, the finite mixture representation allows efficient sampling in the second stage compared with using the raw Monte Carlo posterior draws directly. 
This strikes a balance between fidelity to the first-stage posterior and computational tractability, making it particularly suitable for multi-household data with heterogeneous within-host dynamics.

We fit the mixture model as follows. To simplify notation, we drop the subscripts throughout the remainder of this section, noting that this approximation is constructed at the individual level (i.e. for each individual $i$ within a household $h$).
For a scalar parameter $\theta_j$, define the function
$$
f_1(\theta_j) = \frac{\theta_j - \tilde{m}_{\theta_j}}{\tilde{\sigma}_{\theta_j}}
$$
where $\tilde{m}_{\theta_j}$ and $\tilde{\sigma}_{\theta_j}$ are the sample mean and standard deviation of the $\theta_j$ samples from the within-host posterior. 
For log-transformed parameters, $\tilde{m}_{\theta_j}$ and $\tilde{\sigma}_{\theta_j}$ are the sample mean and standard deviation of the log-transformed samples.
Note that both $\tilde{m}_{\theta_j}$ and $\tilde{\sigma}_{\theta_j}$ are deterministic during the training stage given a set of posterior samples. 
Now, define the function 
$$
f_2(R_0, \delta, \rho, \tau) = \left( f_1(\log R_0), f_1(\log \delta), f_1(\log \rho), f_1(\tau) \right).
$$
To handle non-negativity of $R_0, \delta$ and $\rho$, we fit the GMM to the standardised log-parameters under $f_2$
\[
\boldsymbol{x}^{(W, n)}
= f_2\left(R_0^{(W, n)}, \delta^{(W, n)}, \rho^{(W, n)}, \tau^{(W, n)} \right), \textrm{ for } n = 1, \ldots, n_s.
\]
Fitting a Gaussian mixture to $\{ \boldsymbol{x}^{(W, n)} \}_{n = 1}^{n_s}$ yields an approximation to $p(\boldsymbol{x}^{(W)})$. 
The mixture parameters are estimated by maximum likelihood using the expectation–maximisation (EM) algorithm \citep{reynoldsGaussianMixtureModels2009}.
We use the Julia package \texttt{GaussianMixtures.jl} \citep{leeuwenGaussianMixturesjl2026} to automate the fitting of the GMMs and specify a maximum number of Gaussians of $2^3 = 8$, chosen based on preliminary testing, which indicated this was sufficient to capture the observed distributions.
The fitting process runs the EM algorithm after splitting Gaussians (starting with a single one) to determine the optimal number of components.
Once the GMM has been fitted, we can obtain the density for the raw parameters per individual by applying the change of variables formula to Eq.~\eqref{eq:within_host_posterior_gmm_target},
\begin{equation}\label{eq:within_host_surrogate}
\begin{aligned}
    &\hat{p}\left(\boldsymbol{\theta}^{(W)}, \tau^{(W)} \right) \\ 
    &= \left| \det J_{f_2}\left(\boldsymbol{\theta}^{(W)}, \tau^{(W)}\right) \right| 
   \hat{p}_{\mathrm{GMM}}\!\left( f_2\left( R_{0}^{(W)}, \delta^{(W)}, \rho^{(W)}, \tau^{(W)}\right) \right)
\end{aligned}
\end{equation}
where $J_{f_2}$ is the Jacobian of the transformation and $\hat{p}_{\mathrm{GMM}}(\cdot)$ is the trained GMM distribution. 
As the scalings $\tilde{m}_a$ and $\tilde{\sigma}_a$ are fixed per parameter at the training stage, the determinant of the Jacobian is given by (see Appendix~\ref{app:jacobian_derivation} for derivation)
\begin{equation*}
\begin{aligned}
     &\left| \det J_{f_2}\left(\boldsymbol{\theta}^{(W)}, 
\tau^{(W)}\right) \right|  \\
&= \left(\tilde{\sigma}_{\log R_{0}^{(W)}}\tilde{\sigma}_{\log\delta^{(W)}} \tilde{\sigma}_{\log\rho^{(W)}} \tilde{\sigma}_{\tau^{(W)}}R_{0}^{(W)} \delta^{(W)} \rho^{(W)}\right)^{-1}. 
\end{aligned}
\end{equation*}
Eq.~\eqref{eq:within_host_surrogate} is then used as the individualised prior in Eq.~\eqref{eq:between_host_target}, propagating within-host uncertainty directly into the between-host inference.

\subsection{Inference method details}\label{sec:inference_method}

As mentioned in Section~\ref{sec:within-host_inference}, we sample from the posterior distribution of the within-host model using the MCMC routines described in \citet{morrisRandomTimeshiftApproximation2025a}. 
In brief, this method utilises a Metropolis-within-Gibbs sampler parallelised across four high-performance cores.
The between-host model considered here adopts a similar Metropolis-within-Gibbs structure, but full details are deferred to Appendix~\ref{app:inference_method_details}.
In contrast to the within-host model, we found that, for the between-host model, proposal steps and likelihood evaluations were more efficient when run in serial, rather than parallelised as in \citet{morrisRandomTimeshiftApproximation2025a}.
We do, however, employ parallelism to run three independent Markov chains.
We run three independent chains rather than the typical four due to the memory demands of storing posterior samples across the large number of model parameters and multiple model runs; three chains were sufficient to assess convergence via standard diagnostics.

We employ simple random-walk Metropolis proposals, with parameters blocked into household-level and transmission-level components. 
Local proposal distributions are tuned using a pilot run and then fixed, targeting an acceptance rate of approximately $0.25$ for household-level blocks (12--36 dimensional blocks), and approximately $0.35$ for the low-dimensional (3 dimensional) transmission-related hyper-parameter block, consistent with theoretical and empirical optimal-scaling results \citep{gelmanWeakConvergenceOptimal1997,bedardOptimalAcceptanceRates2008}.

\section{Results}\label{sec:results}

This section presents results evaluating the accuracy, robustness, and computational performance of the proposed inference framework on simulated data. 
We begin by outlining the parameter choices, then examine parameter recovery as a function of the number of households in a dataset. 
Predictive performance is assessed using posterior predictive checks of reported symptom onset times, followed by an analysis of robustness to reduced viral load sampling within households. 
A summary of the computational performance is presented in Appendix~\ref{app:performance}.

\subsection{Simulation studies}\label{sec:simulation_studies}

To validate the proposed inference method, we generated synthetic datasets comprising household outbreaks simulated under known parameter values (see Appendix~\ref{sec:simulation} for full details of the simulation procedure). 
We simulate datasets of $H = 100$ households, with each outbreak generated independently of the other households.
To assess the robustness to stochastic variability in the data-generating process, we simulated 10 independent dataset replicates using identical true parameter values but independent realisations of both within-host and between-host processes.
Household sizes were sampled independently from a modified version of the 2021 distribution of Australian household sizes \citep{householdid} with modifications to omit single person households and pool those of size $N \geq 6$ into the $N = 6$ category.

The true parameter values used to generate all datasets are listed in Table~\ref{tbl:true_parameters}. Within-host parameters were chosen based on previous work \citep{morrisRandomTimeshiftApproximation2025a}, which are consistent with estimated parameters for COVID-19. 
Between-host transmission parameters were selected to yield household secondary attack rates (HSAR)---the proportion of household contacts who become infected, excluding the index case---ranging from approximately 27\% (larger households) to 58\% (smaller households), with an overall HSAR of approximately 40.4\%. 
These HSAR values were chosen to produce realistic heterogeneity across household sizes rather than to match a specific disease, though the estimate of 40.4\% is consistent with HSAR values estimated for the Omicron SARS-CoV-2 variant of 42.7\% (95\% CI $(35.5, 50.4)$) \citep{madewellHouseholdSecondaryAttack2022}.

\begin{table}[!ht]
    \centering
    \renewcommand{\arraystretch}{1.3}
    \begin{tabular}{l|cccccccccc}
        \hline
        \textbf{Parameter} 
        & $\mu_{R_0}$ 
        & $\sigma_{R_0}$ 
        & $\mu_\delta$ 
        & $\sigma_\delta$ 
        & $\mu_\rho$ 
        & $\sigma_\rho$ 
        & $\kappa$ 
        & $\mu_{\omega}$ 
        & $\sigma_{\omega}$ 
        & $\eta$ \\
        \hline
        \textbf{Value}
        & $8$ 
        & $0.5$ 
        & $1.3$ 
        & $0.15$ 
        & $3$ 
        & $0.25$ 
        & $0.5$ 
        & $4.0$ 
        & $0.25$ 
        & $0.1$ \\
        \hline
    \end{tabular}
    \caption{The values of the parameters used to generate the synthetic datasets.}
    \label{tbl:true_parameters}
\end{table}

We initially operate under a paradigm where viral load measurements are assumed to be observed daily for all infected individuals. 
This setting allows validation of the inference procedure under data-rich conditions. 
We later relax this assumption to investigate performance under more realistic, sparser, within-host sampling regimes and delayed household recruitment.
To assess the effect of sample size on inference, we varied the number of households included in the analysis, considering $H \in \{10, 20, 50, 100\}$, which leads to an expected number of 17, 34, 89, and 179 individuals in total to fit to the within host model. 
Since the within-host model has been validated previously \citep{morrisRandomTimeshiftApproximation2025a}, we focus on validating the between-host transmission component and provide only a summary of results for the within-host model here.
For a well-calibrated model, we expect posterior estimates to concentrate around the true parameter values as $H$ increases, with uncertainty decreasing accordingly. Despite using identical true parameters across datasets, stochastic variability at both the within-host and between-host scales induces variation between dataset realisations.

All simulations and inference were performed in Julia v1.12 \citep{bezansonJuliaFreshApproach2017} on a 2021 MacBook Pro with an Apple M1 processor and 16\,GB of memory. 
Pilot runs were used to tune the proposal distributions for both household-specific and shared parameters. 
Following this tuning phase, three independent Markov chains were initialised using posterior means obtained from the pilot runs.
For each dataset, the within-host stage was run for 100{,}000 iterations, discarding the first 20{,}000 as burn-in. The between-host stage was run for 200{,}000 iterations (again discarding the first 20{,}000 iterations), reflecting the increased posterior complexity induced by the household transmission structure.
Run times for each stage of the model and associated effective samples sizes are provided in Appendix~\ref{app:performance}.

Convergence was assessed using a combination of visual diagnostics and formal summary statistics. 
Trace plots for all shared parameters and a representative subset of household-level parameters showed no evidence of poor mixing. 
All parameters had $\hat{R}$ values close to one, indicating satisfactory within- and between-chain convergence \citep{vehtariRankNormalizationFoldingLocalization2021}, and effective sample sizes exceeded 500 in all cases. 
Acceptance rates for the random-walk Metropolis updates were approximately 0.30 for both household-level (averaged across household sizes) and shared parameters, consistent with well-tuned proposal distributions \citep{bedardOptimalAcceptanceRates2008}.

\subsection{Parameter recovery}\label{sec:parameter_recovery}

\begin{figure}[!ht]
    \centering
    \begin{adjustwidth}{-0.5in}{0in}
    \includegraphics{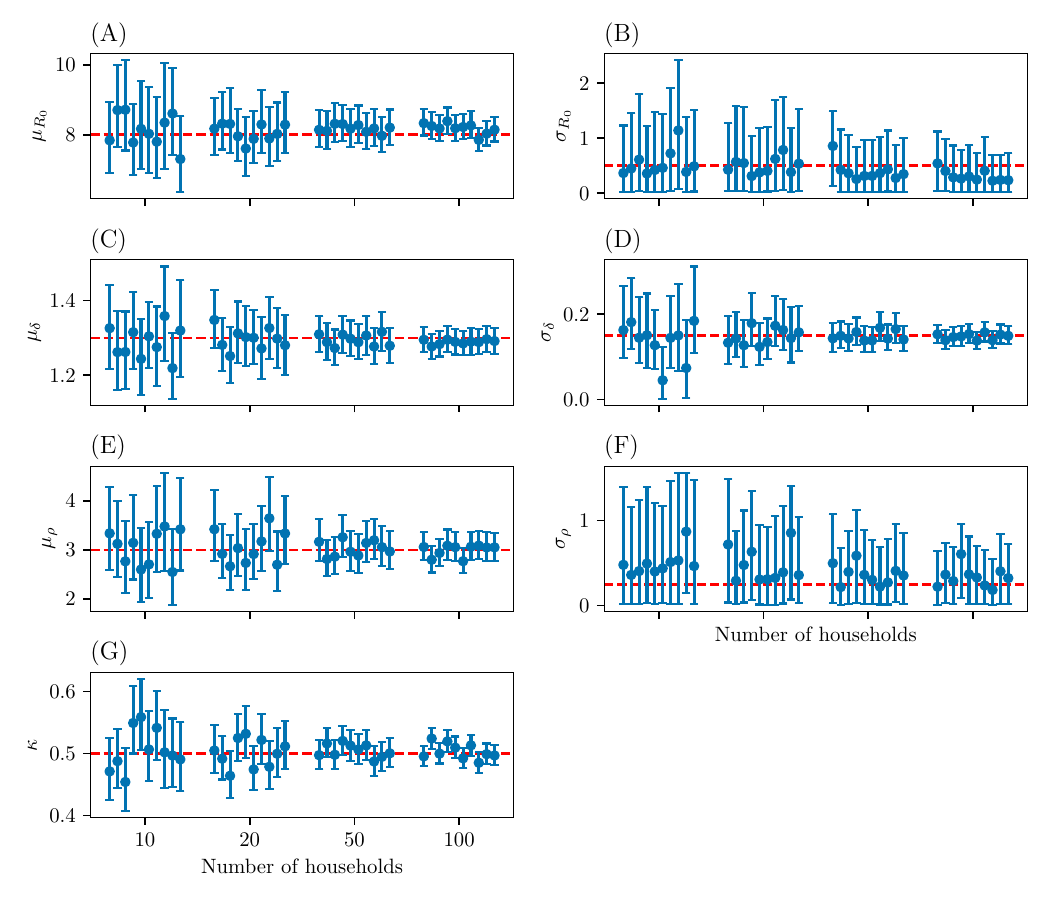}
    \end{adjustwidth}
    \caption{
        Posterior summaries for the within-host parameters, shown as 95\% credible intervals and grouped by number of households in analysis. Each point represents the posterior estimate for that parameter for a data replication with a specified number of households (x-axis).
        In both panels, dashed red lines indicate the true parameter values used to simulate the datasets.
    }
    \label{fig:within-host_cis_by_household}
\end{figure}

We ran the complete inference pipeline described in Section~\ref{sec:inference_method} on each dataset and parameter recovery results are provided here. 
We first examine recovery of the within-host parameters. 
Fig.~\ref{fig:within-host_cis_by_household} presents the 95\% credible intervals for the hierarchical within-host parameters and the measurement noise parameter $\kappa$. 
As the number of households increases, the credible intervals for the population-level mean parameters ($\mu_{R_0}, \mu_{\delta},$ and $\mu_\rho$) contract, and the posterior medians approach the true values used to generate the data, indicating that the method can recover the true parameters when data are plentiful. 
In Fig.~\ref{fig:within-host_cis_by_household}B and D, the scale parameters $\sigma_{R_0}$ and $\sigma_{\delta}$ are well estimated as the number of households grows. 
However, in Fig.~\ref{fig:within-host_cis_by_household}F, $\sigma_\rho$ remains only weakly identified even as the number of households increases.  
From Fig.~\ref{fig:within-host_cis_by_household}G, the posterior distributions for the measurement noise parameter $\kappa$ are centred near the true value, although there are instances of poor coverage (e.g., $\textrm{ID} = 2$, where the true value lies outside the 95\% credible interval).

\begin{figure}[!ht]
    \centering
    \includegraphics{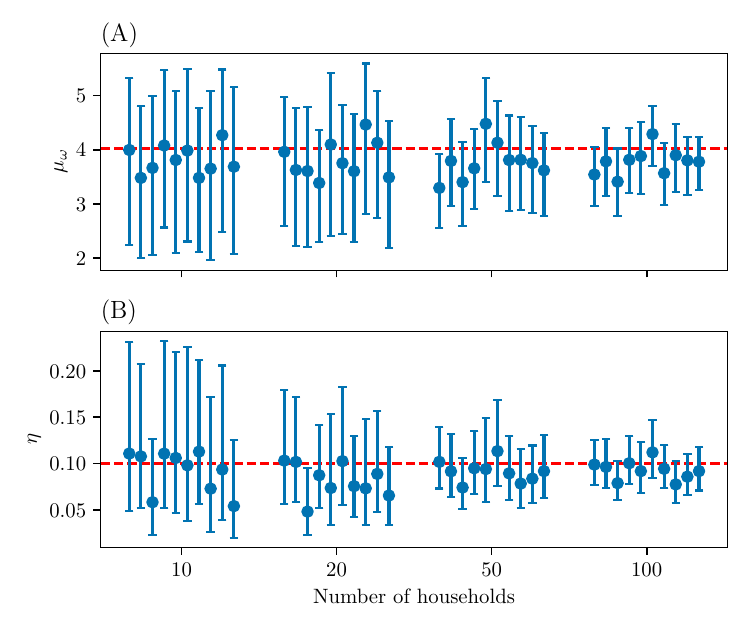}
    \caption{
        Posterior summaries for the between-host parameters for each data replicate, shown as 95\% credible intervals and grouped by number of households in analysis.
        Panel (A) shows $\mu_{\omega}$, while panel (B) shows $\eta$.
        In both panels, dashed red lines indicate the true parameter values used to simulate the datasets.
    }
    \label{fig:cis_by_household}
\end{figure}

Fig.~\ref{fig:cis_by_household} presents the 95\% credible intervals for the between-host parameters $\mu_{\omega}$ and $\eta$. 
The standard deviation of the log-10 half saturation thresholds, $\sigma_{\omega}$, is omitted, as it is not identifiable and remains largely prior-determined even with a large number of households.
As the number of households increases, the credible intervals for $\mu_{\omega}$ and $\eta$ contract, and the posterior medians approach the true values used to generate the data, indicating that the method can recover the true parameters when data are plentiful.
The convergence behaviour as the number of households increases is also consistent across the dataset replicates, suggesting that the method is well calibrated. 
There appears to be a slight negative bias in $\mu_{\omega}$, likely arising from limitations in the within-host component. In particular, viral load measurements are primarily observed above the detection limit ($\nu(t) > 2.65$), while observations below this threshold contribute only through censoring, reducing information about the early growth phase. Additionally, slight overestimation of $R_0$ in some datasets, through its relationship with the growth rate \citep{maEstimatingEpidemicExponential2020}, can further contribute to mischaracterisation of early trajectory dynamics. 
Together, these effects lead to weak identification of the peak timing and early growth rate of the viral trajectory and a tendency for inferred trajectories to exhibit earlier peak times. Since $\mu_{\omega}$ governs the viral load threshold at which symptom onset occurs, earlier inferred trajectories cross this threshold sooner, and to remain consistent with observed onset times, the model compensates by favouring lower values of $\mu_{\omega}$.

\subsection{Predictive distributions}\label{sec:pred_distributions}

We assess model fit via posterior predictive distributions for the observed symptom onset time assuming infection at time $0$, which we denote $d^\star$.
The simulated data is converted to this scale by computing $d_{h, i} - u_{h, i}$ (using the known infection times $u_{h, i}$) for each individual $i$ in household $h$.
These times may be derived for specific individuals or household configurations, or considered in aggregate as draws from the population-level distribution over infectious individuals; the latter perspective provides a clearer assessment of predictive performance.

\begin{figure}[!ht]
    \centering
    \includegraphics{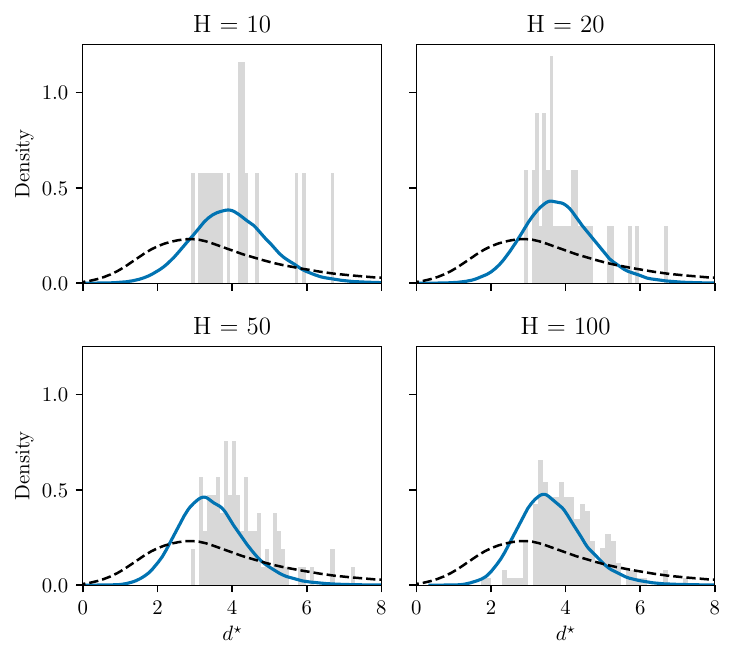}
    \caption{
        Predictive distributions for $d^\star$ by number of households (shown as panel titles). 
        The prior predictive is indicated by the dashed black line, while blue lines correspond to posterior predictive distributions for different dataset sizes.
        The observed data used in the respective models is shown as a grey histogram. 
    }
    \label{fig:ppc}
\end{figure}

Samples from both the prior and posterior predictive distributions are obtained by sequentially sampling from the model (see Appendix~\ref{app:predictive_distributions} for details), and then  
symptom onset times are computed as described in Section~\ref{sec:data}. 
Fig.~\ref{fig:ppc} shows predictive distributions for $d^\star$ for one of our simulated datasets ($\textrm{ID} = 1$). 
We observe that even with few households ($H = 10, 20$), $d^\star$ is well inferred, matching that of the larger dataset sizes ($H = 50, 100$), with the posterior predictive distribution shifted away from the prior and exhibiting a reduced upper tail. 
Additional data yields only modest further reductions in uncertainty, resulting in slightly narrower distributions as $H$ increases. 
The posterior predictive distribution is centred slightly earlier than the observed data, consistent with the negative bias in $\mu_{\omega}$ (Section~\ref{sec:parameter_recovery}), which shifts inferred trajectories to cross the onset threshold earlier. 
These results indicate that the model provides reasonable predictive performance even with limited data, with only modest gains from increasing sample size.

\begin{figure}[!ht]
    \centering
    \includegraphics{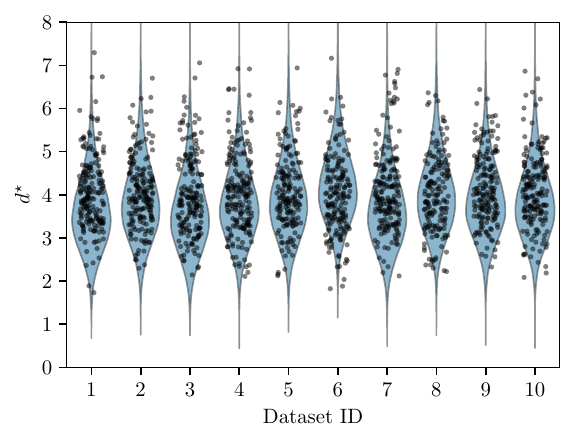}
    \caption{
        Predictive distributions for $d^\star$ (for $H = 100$ households) shown as violin plots for each dataset. Black dots indicate the observed $d^\star$ and are jittered horizontally and partially transparent to reduce overplotting and provide a visual approximation to the distribution.
    }
    \label{fig:ppc_with_data}
\end{figure}

Fig.~\ref{fig:ppc_with_data} shows the posterior predictive distribution of $d^\star$ for the $H = 100$ households, for each dataset, overlaid with the observed $d^\star$.
Overall, the model captures the central tendency and tail behaviour of the observed distributions, with strong agreement between the posterior predictive distributions and the observed data (points).

\subsection{Reduced frequency within-host data measurements}

To assess model performance under more realistic observational conditions, we consider a dataset in which the within-host data are post-processed to reflect a more plausible household study design with reduced sampling frequency \citep{FirstFewCases2021}.
We generate the underlying data as described in Section~\ref{sec:simulation_studies}, and subsequently apply additional processing steps to the simulated within-host observations.
For simplicity, we assume $H = 50$ households, treating the daily observation scenario as the reference ``truth''.
We have intentionally not shown the observed symptom onset times in the figures, focusing here on the effect of reduced viral load sampling on posterior estimates.

To generate this data we assume a random recruitment time for the household occurring 0--3 days after infection of the index case, consistent with a scenario in which individuals are identified as close contacts of a known infectious case. 
Following recruitment, viral load measurements are assumed to be collected every $r$ days rather than daily, substantially reducing the temporal resolution of the observed within-host trajectories.
All other aspects of the data-generating process remain unchanged, including the onset-time observations, which are assumed to be accurately recorded via participant symptom diaries.

We consider three reduced-frequency cases in addition to the reference daily scenario: three-day sampling, five-day sampling, and weekly sampling. 
The three-day sampling reduces the number of observations by two-thirds but is still frequent enough to capture key viral events. 
The five-day and weekly scenarios are more realistic and result in more variance in the within-host dynamics.

\begin{figure}[!ht]
    \centering
    \begin{adjustwidth}{-0.5in}{0in}
        \includegraphics{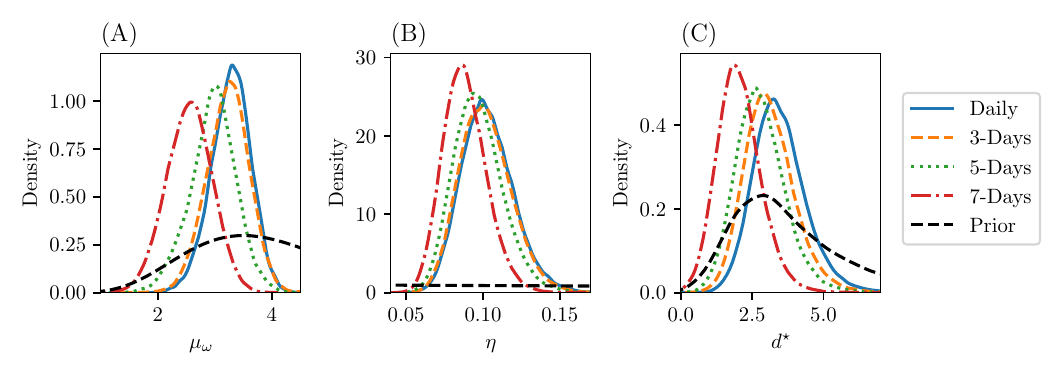}
    \end{adjustwidth}
    \caption{ 
        (A, B) Marginal posterior distributions for the transmission parameters, $\mu_\omega$ and $\eta$, under varying viral load sampling regimes. 
        (C) Posterior predictive $d^\star$ for a newly sampled individual under varying viral load sampling regimes.
        Coloured lines correspond to distributions for the different regimes.
        The reference case with daily sampling is shown by the solid blue line. 
    }
    \label{fig:censored_posteriors}
\end{figure}

Fig.~\ref{fig:censored_posteriors}A and B show the marginal posterior distributions for $\mu_{\omega}$ and $\eta$ under the different sampling regimes, with daily sampling treated as the truth. 
There is minimal difference between the daily and three-day regimes, with distributions appearing very similar. 
The five-day sampling regime results in modest downward shifts in $\mu_{\omega}$ and $\eta$, although the change in posterior for $\eta$ is much smaller, remaining  similar to the daily regime. 
This is expected as $\eta$ is primarily a function of the final size (the total number of individuals infected in the household), and is readily identified given sufficient viral load trajectories and number of individuals displaying symptoms. 
With a 7-day sampling scheme, differences become more pronounced. 
The posterior for $\eta$ remains close to the daily sampling result, consistent with its dependence on final size. 
In contrast, the posterior for $\mu_{\omega}$ is substantially lower, attaining a median of approximately $2.5$ compared with $3.4$ for daily sampling.
Fig.~\ref{fig:censored_posteriors}C shows the downstream effect on posterior predictive $d^\star$ for the same regimes. 
As the frequency, and hence number, of viral load measurements decreases, $d^\star$ is increasingly underestimated relative to the daily sampling. 
Once more, minimal difference is observed between the daily and three-day sampling regimes, reinforcing that frequent testing, but not necessarily daily, captures the key dynamics necessary for accurate inference.

To view the effect of these differences in inferred posteriors and how they translate into the household level dynamics we can look at a specific household realisation.
We present an illustrative example from one dataset, showing the true infection times alongside the posterior predictive distributions for each individual. 
Although this example does not fully characterise the model's capabilities across all household configurations, it demonstrates the model’s capacity to recover plausible transmission chains and highlights the effects of the reduced sampling frequency. 

\begin{figure}[!ht]
    \centering
    \includegraphics{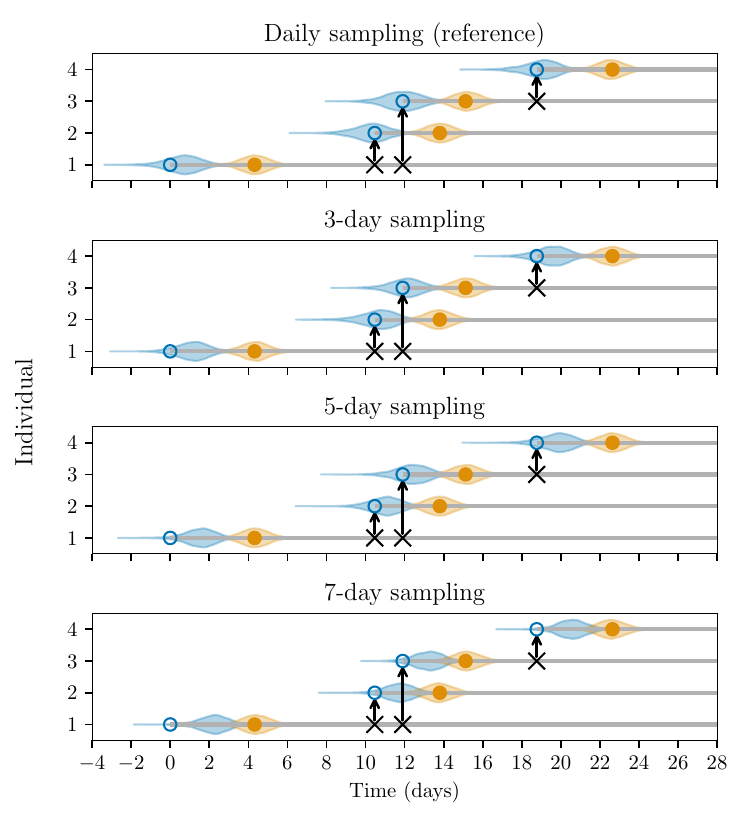}
    \caption{
        Inference results for an outbreak in a household of four individuals under the various sampling regimes.
        Posterior predictive distributions of infection times and reported symptom onset times are shown as blue and orange violin plots, respectively, corresponding to kernel density estimates. 
        Grey horizontal lines indicate where an individual is infected (but not necessarily infectious). 
        Open blue dots represent unobserved infection times, and solid orange dots indicate reported symptom onset times. 
        Black crosses mark transmission events, with vertical arrows connecting inferred infector–infectee pairs (note, this is unobserved).
        The title of each panel gives the censoring regime. 
    }
    \label{fig:household8_line_list_predictions}
\end{figure}

Fig.~\ref{fig:household8_line_list_predictions} presents inference results for a household of four individuals. 
True (unobserved) infection times are indicated by open blue points, and reported symptom onset times by orange points.
Posterior predictive distributions for infection and symptom onset times are visualised using kernel density estimates, shown as violin plots and colour-coded by event type. 
Infection times and transmission events are unobserved, and the inference relies solely on the symptom onset data and viral load measurements. 
The method's ability to recover the infection times and transmission parameters from such indirect data is therefore encouraging. 
In the case of the daily sampling regime, the true values fall within the inferred distributions, indicating accurate recovery of both times.
As the viral load sampling becomes less frequent, we see that the model biases inference by matching the reported symptom onset time and the infection times become biased towards the onset time. 

\begin{figure}[!ht]
    \centering
    \includegraphics{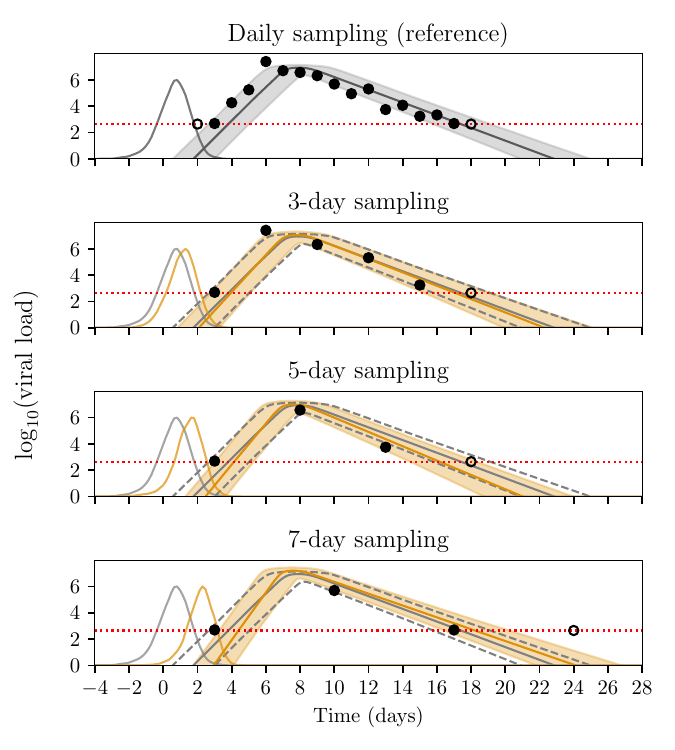}
    \caption{
        Posterior predictive viral load trajectories for a single individual in the example dataset under the different sampling regimes. 
        Orange lines correspond to the specific sampling regime and grey lines correspond to the daily sampling as a reference.
        The daily sampling is given in the top panel for ease of data comparison. 
        The non-daily sampling regimes have the 95\% prediction intervals shown in orange and the daily sampling is not shaded, but is instead indicated with dashed lines. 
        The row indicates the sampling regime, indicated on the left-hand side.
        The marginal posterior for the infection times are indicated by kernel density estimates near the start of the trajectories, colour coded to the relevant case. 
        Black dots indicate the observed viral loads (after processing). Solid dots are those above the detection limit and open dots are those at or below the limit. 
        Differences in displayed data arise due to processing as outlined at the start of Section~\ref{sec:within-host_inference}.
    }
    \label{fig:household8_vl_predictions}
\end{figure}

Fig.~\ref{fig:household8_vl_predictions} shows the posterior predictive viral loads under the different sampling regimes for a representative individual. 
Examining Fig.~\ref{fig:household8_vl_predictions} together with Fig.~\ref{fig:household8_line_list_predictions} reveals how the sampling regimes shape the marginal posterior for $\mu_{\omega}$ and posterior predictive $d^\star$.
For this household, the recruitment time was day 3, which only applies to the reduced sampling (3, 5, and 7 day) instances.
Note that other individuals exhibit similar characteristics and are shown (for this household) in Appendix~\ref{app:predictive_vl_distributions}. 
Overlaid are the infection time marginal posteriors for the individual. 
The reference case of daily sampling is given in grey and is shown in the top panel. 
Note that observed data appear different from the daily measurements due to the processing of the time series. 
As described in Section~\ref{sec:within-host_inference}, viral load data are centered with the peak measurement at $t = 0$ and truncated to 3 consecutive limit of detection observations.
Even under very infrequent sampling (7-days), the within-host model reasonably captures the growth and decline phases. 
However, as the sampling frequency reduces, we see that the log-viral load tends to bias the growth later and have a faster growth rate. 
Under sparser sampling, the window of plausible peak times widens and the model tends to infer the peak towards the centre of this window, resulting in an earlier inferred peak time. 
In addition, with less viral load data the model prefers later infection times. 
These two effects couple to explain the lower $\mu_{\omega}$ estimates relative to daily sampling, and subsequently, $d^\star$ is also shorter. 
This suggests that frequent sampling is crucial to capture the early stages of infection and enable accurate infection time inference.

\subsection{Using out of study data to inform within-host dynamics}\label{sec:oos_data}

With the model performance under different sampling regimes demonstrated, we turn our attention to an advantage of our cut approach to inference. 
The main limitation in the reduced sampling regime is that the viral load dynamics are misspecified due to growth and decline being captured incorrectly. 
This is unavoidable as the data simply does not describe the behaviour in these regimes. 
We can improve this situation through the inclusion of out-of-study (OOS) data to inform the within-host dynamics as this model is decoupled from the between-host model.

Suppose in addition to the household dataset, $\{\mathcal{Y}_h, \boldsymbol{d}_h\}_{h \in \mathcal{H}}$, we also have viral load data, $\mathcal{Z}$, from some other study. 
This data could come from bubble arrangements in sports---where staff and players travel in isolation to the general public and are routinely tested---as happened during the COVID-19 pandemic \citep{kisslerstephenm.ViralDynamicsSARSCoV22021}, or from other sources \citep{clearyUsingViralLoad2021}.
The main purpose of including such data is that it may contain more information than just using the low-frequency sampled viral loads from the households.
Assuming the same within-host model applies to each dataset, we can fit to the combined dataset $\{\{\mathcal{Y}_h\}_{h \in \mathcal{H}}, \mathcal{Z}\}$ to obtain a better within-host fit through the additional precision gained from tighter estimation of the shared hyper-parameters. 

We explore how the inclusion of additional data influences household estimates by fitting the model for $H = 50$ households in the ``worst" situation of only 7-day testing. 
The additional dataset consists of another 90 individuals (obtained by simulating another 50 households) where the viral load data is taken at daily resolution. 
These individuals are only used to fit the within-host model and are not included in the between-host model stage of fitting.

\begin{figure}[!ht]
    \centering
    \begin{adjustwidth}{-0.5in}{0in}
        \includegraphics{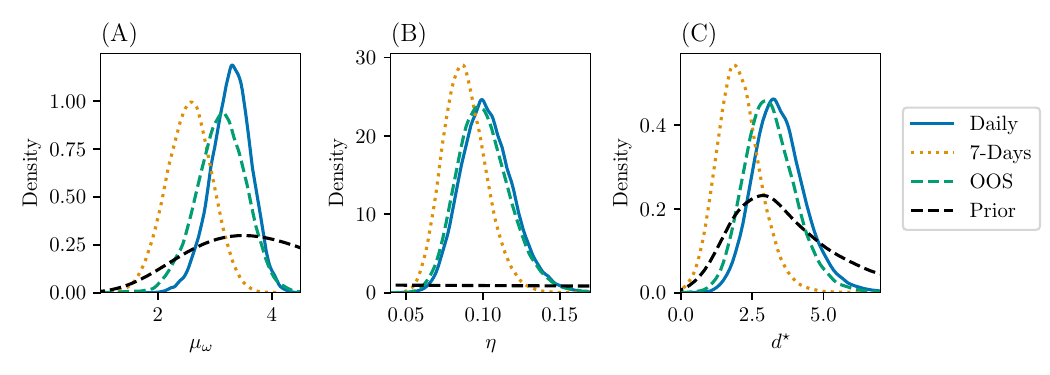}
    \end{adjustwidth}
    \caption{ 
        (A, B) Marginal posterior distributions for the transmission parameters under the different within-host data regimes. 
        (C) Posterior predictive distributions for $d^\star$ under varying viral load sampling regimes. 
        Coloured lines correspond to distributions for the different regimes.
        The reference case is shown by the solid blue line. 
        OOS refers to out-of-sample fit. 
    }
    \label{fig:oos_posteriors}
\end{figure}

Fig.~\ref{fig:oos_posteriors}A and B shows the marginal posteriors for the transmission parameters under daily (solid line), weekly (dotted line), and weekly with included OOS data (dashed line) experimental setups. 
We can see that the inclusion of the OOS within-host data substantially reduces the underestimation of the transmission parameters, with the blue dotted lines lying closer to the daily sampling. 
There is slight downwards bias in $\mu_{\omega}$ but this is reduced considerably compared to no OOS data. 
The effect from including OOS data on the posterior predictive $d^\star$ values are shown in Fig.~\ref{fig:oos_posteriors}C. 
We see that the marginal distribution for the OOS case lies much closer to the daily sampling case. 
This suggests that the inclusion of external data can correct for the reduced testing frequency.

\section{Discussion}\label{sec:discussion}

This work presents a framework for linking within-host and between-host infection dynamics. 
We do so by constructing a force of infection function for each individual, driven by their underlying viral load. This captures how an individual's capacity to transmit varies over the course of their infection, which in turn drives a generation time epidemic model for the between-host transmission dynamics.
Our approach provides a pragmatic, yet accurate, means of connecting the two model scales, enabling information to flow coherently between individual-level infection processes and population-level spread.

We show how this methodology can be used for both forward simulation and statistical inference in a simulated experimental setup of a household transmission study. The results demonstrate that the method can reliably estimate key between-host model parameters as the amount of available data increases and when the sampling of viral loads is frequent enough. 
Posterior credible intervals for the mean population viral load threshold, $\mu_{\omega}$, and the transmission rate, $\eta$, contracted systematically with increasing numbers of observed households, with posterior medians converging toward the true values used to generate the data. 
This behaviour indicates that increasing the number of observed households drives improved parameter recovery; however, inference performance is contingent on adequate within-household resolution, and poorly resolved households contribute little information to the likelihood. This pattern was observed consistently across independently simulated datasets, suggesting robustness to stochastic variation in the within- and between-host processes.

With frequent viral load measurements, posterior uncertainty was substantial at smaller sample sizes, with some variability between dataset replicates. This included a systematic deviation in one replicate, likely reflecting an atypically sparse transmission chain by chance.
This is expected in settings with limited information (i.e.\ few households with limited transmission) and reflects the combined effects of stochastic transmission dynamics and partial observations.
As the number of households increased, these discrepancies diminished, and posterior summaries stabilised across replicates. 
In contrast, the standard deviation of the viral load threshold distribution, $\sigma_{\omega}$, was not identified by the data and its posterior remained close to the prior across all sample sizes, highlighting a structural limitation of the available data to inform this parameter.
Household-level predictive checks show that the model can accurately infer individual infection and symptom onset times, with posterior predictive distributions generally centering on the true values for each household member.
This provides complementary evidence to the between-host parameter recovery results above, together demonstrating that the approach can resolve both aggregate transmission dynamics and individual-level infection trajectories.

We also examined model performance under reduced viral load testing frequency. 
Inference remained largely consistent when observations were collected every 1–3 days, but further reductions to 5-day or weekly sampling degraded parameter estimation and shifted posterior predictive infection times earlier. 
As noted in Section~\ref{sec:oos_data}, this occurs because fewer informative viral load observations weaken identification of the growth phase, leading the model to infer trajectories that rise earlier than when sampling is daily. 
Notably, even under these lower-frequency sampling regimes, the model reliably recovered the infection rate $\eta$, consistent with its dependence on the outbreak final size rather than detailed viral load trajectories. 
Together, these results demonstrate that the proposed approach can recover identifiable between-host parameters in realistic household study designs, while highlighting the importance of adequate viral load sampling for accurately estimating individual-level dynamics and associated transmission patterns.

Our approach is explicitly hierarchical at both the within- and between-host levels.
One of the benefits of this approach is that it allows the within-host parameters to be estimated from larger datasets of viral loads that may not be household-based, including data from studies outside the current household cohort. 
The hierarchical structure of the within-host model enables more informative trajectories to guide population-level parameter estimates, improving inference even when some individuals have low-frequency sampling.
As shown in Section~\ref{sec:oos_data}, incorporating out-of-sample data can compensate for low frequency household observations, yielding estimates comparable to those obtained from the daily-resolution data.
This has potential implications for how household studies are conducted. 
For instance, a subset of households could have increased monitoring, with more frequent testing, but the majority could be sampled in a sparser way. 
This would reduce the financial burden of the study while still offering robust parameter estimates. 

Parameter inference is a pivotal component of effective modelling for informing public health measures during emerging epidemics \citep{marionModellingUnderstandingPandemics2022,swallowChallengesEstimationUncertainty2022}. 
However, inference in multiscale models is non-trivial, and it is therefore unsurprising that the existing literature has largely focused on forward simulation \citep{almoceraMultiscaleModelWithinhost2018,goyalMultiscaleModellingReveals2022,smithEfficientCouplingWithinand2025,yinAccurateStochasticSimulation2025}. 
These approaches have made important progress in bridging within- and between-host scales, though they typically make the simplifying assumption of a deterministic within-host model that is treated as identical across infected individuals, with stochasticity arising only at the between-host scale. 
Our work addresses the open problem of inference in this multiscale setting; we not only outline the forward simulation, but also the inverse problem of parameter estimation using a model that is hierarchical at both the within- and between-host levels. This approach relaxes the assumption that within-host dynamics are identical across infected individuals and explicitly captures stochastic variability arising from the underlying process. Our use of a random time shift model for the within-host dynamics is the key ingredient that makes inference tractable. 

The methods detailed here serve as an initial realisation of the way the within- and between-host models can be coupled but is by no means exhaustive. 
Our within-host model and the resulting intensity function are both simplistic and as such are missing features that may be important, such as immune responses or complex infection dynamics.  
Additionally, the way symptom onset timing and the within-host dynamics are linked (via a hill function) is also simplistic---even though it is consistent with how this is often implemented using compartmental models \citep{keelingModelingInfectiousDiseases2008}.
Nevertheless, the approaches outlined here can easily accommodate more complex viral models and intensity functions. 
A natural change could be the inclusion of a more mechanistically detailed within-host model incorporating immune dynamics \citep{challengerModellingUpperRespiratory2022,zitzmannHowRobustAre2024}, or a force of infection function that is proportional to the viral load or accounts for things like the removal of infected individuals or interventions like antivirals \citep{ashcroftTesttraceisolatequarantineTTIQIntervention2022}. 
Additionally, one could include complexity of pre-symptomatic infectiousness in the intensity function through the addition of another threshold parameter. 
Combined with accurate inference of the time of infection, this could offer insight into the otherwise difficult problem of characterising the relationship between symptom onset and infectiousness \citep{schultzeCOVID19HumanInnate2021,margiottaInvestigatingRelationshipImmune2025}.

A more fundamental reformulation would be to jointly represent viral load and symptom onset via an explicit immune-response component, treating symptom onset as an emergent property of the within-host dynamics rather than an externally specified parameter.
In such a formulation, onset-related parameters would be absorbed into the within-host model and would no longer enter the between-host transmission model directly. 
This does not invalidate our two-stage framework: posterior samples of the relevant latent within-host quantities could still be used as input to the between-host model, preserving the same inference procedure.
The present formulation therefore reflects a pragmatic choice driven by data availability and computational considerations, rather than a limitation of the two-stage inference strategy itself. 
How symptom onset relates to immune response dynamics and disease severity remains an active area of research \citep{handelCrossingScaleWithinhost2015,schultzeCOVID19HumanInnate2021,margiottaInvestigatingRelationshipImmune2025}.

A primary challenge of our framework lies in the complexity introduced by several novel model components, and the potential for implementation and/or approximation error that accompanies them.
A key source of complexity is our reliance on custom MCMC schemes, which require manual tuning to obtain reliable posterior samples, as assessed by diagnostics such as $\hat{R}$ and effective sample size. 
While such challenges are common in models of this class and we have tuned our proposals to achieve reasonable efficiency, substantial speedups may be achievable by leveraging automatic differentiation, for example through probabilistic programming frameworks such as Stan \citep{stan2024} or libraries such as JAX \citep{jax2018github}.
However, doing so would require the model to be amenable to automatic differentiation, which is not currently the case due to several parts that are discontinuous or non-differentiable. For example the optimisation steps embedded within the likelihood evaluation to obtain the mean of the onset distribution is non-differentiable. 
Addressing this would require substantial reformulation of the model.

The models developed here are specifically tailored to household-structured data, where transmission events and exposure histories can be resolved at the individual level and within-host dynamics can be naturally linked to between-host spread. This focus informs both the model structure and the design of the proposal distributions used for inference.
Extending the framework to larger, more weakly structured populations (e.g. hospital wards or boarding schools) with fewer but longer transmission chains, would require a substantial redesign of the proposal mechanism rather than a direct scaling of the current approach. 
This is not the same as including out-of-study data as in Section~\ref{sec:oos_data}. 
We would effectively be fitting a single large ``household'' with many individuals---a structurally different problem.
Such an extension would also introduce computational challenges, as naive implementations would require repeated evaluation of the full transmission model when updating individual-level parameters. In addition, larger population settings are typically subject to greater observation error, including missing data and partial case or viral load detection, which may further limit model applicability. Finally, inference in large populations is often based on a single realisation of the transmission process, which may constrain parameter identifiability relative to household-based analyses. As such, the present formulation is best viewed as a household-level modelling framework, with extension to large population settings requiring new methodological developments rather than a straightforward generalisation.

In conclusion, we have presented a framework for Bayesian inference across within-host and between-host scales of infection dynamics.
By constructing a force of infection function from within-host viral trajectories via random time-shifts, we obtain a probabilistic model structure that is tractable for inference, enabling transmission parameters to be estimated directly from individual viral load and symptom onset data.
Our model and subsequent methods introduced here offer a framework for advancing the study of transmission dynamics during the outbreak of a novel disease or planned longitudinal study.   

\appendix

\begin{appendices}

\section{Example trajectories for hybrid model}\label{sec:example_trajectories}

\begin{figure}[!ht]
    \centering
    \includegraphics{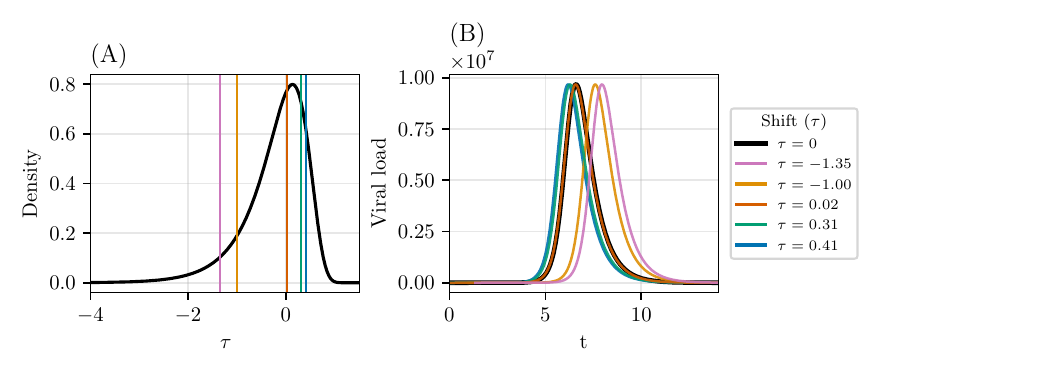}
    \caption{
        Example trajectories of the hybrid model described in Section~\ref{sec:within_host_model} with parameters $\boldsymbol{\theta} = (8.0, 4.0, 1.3, 3.0, 10.0)$ and $u = 0$. 
        (A) shows the time-shift distribution (black) arising from this set of parameters and 5 sampled realisations are indicated by vertical coloured lines. 
        (B) shows the realisations of the viral load through time for these realisations of $\tau$. The black curve shows the deterministic trajectory (i.e. solution to Eq.~\eqref{eq:odes}) with no shift. The coloured lines are matched to their realised $\tau$ values and show the effect of shifting the solution. 
    }
    \label{fig:example_trajectories}
\end{figure}

Fig.~\ref{fig:example_trajectories} shows example behaviour for our hybrid model.
We show the time-shift distribution in Fig.~\ref{fig:example_trajectories}A and indicate realisations from this distribution with vertical coloured lines. 
The trajectories from the hybrid model are shown in Fig.~\ref{fig:example_trajectories}B alongside the raw solution to Eq.~\eqref{eq:odes}. 
The key behaviour is that if $\tau > 0$ (i.e. for the red, green and blue curves) then faster growth results and hence an earlier peak. 
This can be considered equivalent to an earlier infection time (i.e. $u < 0$). 
In the case that $\tau < 0$ (i.e. pink and orange curves) there is delayed growth and hence a later peak. 

\section{Forward simulation of the joint model}\label{sec:simulation}

\subsection{Within-host simulation}

We begin by describing the forward simulation for a single infected individual that can then be used to simulate an outbreak in a population as described in the next section. 
Assume individual $j$ is infected at time $t = u_j$.
First, we draw within-host viral parameters for the individual $\boldsymbol{\theta} = (R_0, \delta, \rho)$ and $\omega$. 
We then sample $\tau$ from the random time-shift distribution given $\boldsymbol{\theta}$. 
Next, we solve for this individual's log-10 viral load, $\nu_j(t)$ and hence their force of infection, $\lambda_j(t)$, given by Eq.~\eqref{eq:lambda}. 
Infection attempts occur as a non-homogeneous Poisson process with intensity $\lambda_j(t)$. 
The cumulative force of infection is given by 
\begin{equation}\label{eq:cumulative_lambda}
    \Lambda_j(t) = \int_{u_j}^{t} \lambda_j(s) \dif{s},
\end{equation}
where $\Lambda_j(\infty)=\lim_{t\rightarrow \infty}\Lambda_j(t)$.
Let $n_{j}$ denote the number of infection attempts the individual makes over the course of their infection, $[0, \infty)$, then 
$n_j \sim \textrm{Poisson}(\Lambda_j(\infty))$.
If $n_j=0$ then there are no attempts, otherwise if $n_j \ge 1$ then the corresponding relative infection time attempts are sampled from \citep{pasupathy2011},
\begin{equation}\label{eq:relative_foi}
    g_{j}(s) = \frac{\lambda_j(s)}{\Lambda_j(\infty)}.
\end{equation}
We can numerically sample from $g_j$ by using the inverse cumulative distribution function (CDF) method, noting that the CDF is given by 
$G_j(s) = \frac{\Lambda_j(s)}{\Lambda_j(\infty)}$. 
Sampling yields an unsorted set of relative infection attempt times which we collect in the vector $\boldsymbol{s}_j$.
The procedure for generating the individual dynamics is summarised in Algorithm \ref{alg:individual_simulator}.

Our assumption that the observations of the viral load from testing do not directly influence the between-host dynamics means we can simulate the viral load observations after the main simulation loop, performing the processes outlined in Section~\ref{sec:data}.
As a part of the individual simulation we also simulate the reported symptom onset time for the individual, $d_j$. 
This is obtained through the approach described in Section~\ref{sec:between_host_model}.

\subsection{Simulation of transmission model}

Let $\boldsymbol{Y}(t)$ be a vector denoting the infection status of the $N$ individuals in the population at time $t$. 
At each time $Y_i(t) = 0$ if individual $i$ is susceptible, otherwise $Y_i(t) = 1$. 
There is no need to distinguish when an individual is recovered as we can assume this is the case when they stop making infection attempts. 
Let $S(t)$ and $I(t)$ denote the total number of susceptible and infectious individuals at time $t$, which can be recovered from the labels
\begin{align*}
    S(t) = \sum_{i = 1}^{N} \mathbb{I}_{\left\{ Y_i(t) = 0 \right\}}, \\ 
    I(t) = \sum_{i = 1}^{N} \mathbb{I}_{\left\{ Y_i(t) = 1 \right\}}. 
\end{align*}
Note that both $S(t)$ and $I(t)$ are not needed to be stored during the simulation and are only used for deciding whether an infection attempt is successful.

We assume an initially infected individual at time $t=0$. Their dynamics are simulated as described above, generating a set of infection attempt times $\mathbf{s}_1$. To track these and infection attempts from subsequently infected individuals we use a priority queue data structure \citep{Cormen2022}, denoted $\Psi$. 
When an individual is infected, we add the possible infection attempt times to $\Psi$, which are automatically sorted.
Next we pop the first elements off the queue, which is the time of the next infection attempt $s_{1, i}$. 
We sample $u \sim \text{Uniform}(0, 1)$ and if $u < S(s_{1, i}) / (N - 1)$, the infection attempt is successful, otherwise nothing happens (the infection attempt was with an already infected individual). 
In the case of a successful attempt, we select an individual with index $j \in \{j : Y_j(s_{1, i}) = 0\}$ and set $Y_j(s_{1, i}) = 1$. 
Now we can solve for this individual's within-host dynamics and simulate the times of their infection attempts, $\boldsymbol{s}_{j}$, adding those times to $\Psi$.
Information about infected individuals is recorded as the simulation proceeds by tracking an event log, $\mathcal{A}$, that stores information for individuals including their index, the time of infection and symptom onset. 
For example, once a new infection occurs, we add the events $(j, \textrm{infection}, t)$ and $(j, \textrm{onset}, d_j)$ to $\mathcal{A}$.

Algorithms~\ref{alg:individual_simulator} and~\ref{alg:population_simulator} describe the main control flow for generating a realisation of a single outbreak within a closed population.
Specifically, Algorithm~\ref{alg:individual_simulator} provides the individual simulator which is responsible for sampling parameters for an individual, simulating their viral load trajectory and generating the infection attempts. 
This algorithm also returns the reported symptom onset time for that individual. 
Algorithm~\ref{alg:population_simulator}, describes how Algorithm~\ref{alg:individual_simulator} can be used to generate an outbreak in a closed system of size $N$ over a time horizon $T_f$. 

\begin{algorithm}[!ht]
\caption{Individual simulator}
\label{alg:individual_simulator}
\begin{algorithmic}[1]\raggedright
    \Statex\hspace{-\algorithmicindent}\textbf{Input:} Individual index $j$, infection time $u_j$, prior distributions $p(\boldsymbol{\theta} \mid \boldsymbol{\phi})$, $p(\omega)$, infection rate $\eta$.
    
    \vspace{0.5em}
    
    \vspace{0.5em}
    \State Sample within-host parameters $\boldsymbol{\theta}_j \sim p(\boldsymbol{\theta} \mid \boldsymbol{\phi})$.
    \State Sample $\omega_j \sim p(\omega)$
    \State Solve the within-host model with parameters $(\boldsymbol{\theta}_j, u_j, \omega_j, \eta)$ to obtain $\nu_j(t)$ and $\lambda_j(t)$.
    \State Solve for reported symptom onset time, $d_j$.
    \State Compute cumulative infectiousness
    \[
    \Lambda_j(t) = \int_{0}^{t} \lambda_j(u)\,du.
    \]
    \State Sample the number of infection attempts
    \[
    n_j \sim \text{Poisson}(\Lambda_j(\infty)).
    \]
    \State If $n_j>0$, sample attempt times $\boldsymbol{s}_{j} = \{s_{j,1},\ldots,s_{j,n_j}\}$ from
    \[
    G_\lambda(s) = \frac{\Lambda_j(s)}{\Lambda_j(\infty)}.
    \]
    
    \State\textbf{Return:} Infection attempt times $\boldsymbol{s}_{j}$, reported symptom onset time $d_j$, and log-10 viral load trajectory $\nu_j(t)$.
\end{algorithmic}
\end{algorithm}

\begin{algorithm}[!ht]
\caption{Population level simulator}
\label{alg:population_simulator}
\begin{algorithmic}[1]\raggedright

    \Statex \hspace{-\algorithmicindent}\textbf{Require:} Population size $N$, time horizon $T_f$, prior distributions $p(\boldsymbol{\theta} \mid \boldsymbol{\phi})$, $p(\omega)$, infection rate $\eta$.
    
    \vspace{0.5em}
    
    \State Initialise the event log $\mathcal{A} = \{\}$, priority queue $\Psi = \{\}$, and viral trajectory storage $\mathcal{B} = \{\}$.
    \State Initialise state of each individual in the system $\boldsymbol{Y}(t) = (1, 0, \ldots, 0)$ where $j = 1$ is the initial infection.
    \State Set $t = 0$.
    \State Run Algorithm~\ref{alg:individual_simulator} for an initial infection with index $j = 1$ at time $t = 0$, returning infection attempt times $\boldsymbol{s}_{1}$, reported symptom onset time $d_1$, and $\nu_1(t)$.
    \State Push $\boldsymbol{s}_{1}$ onto $\Psi$ and store $\nu_1(t)$ in $\mathcal{B}$.
    \State Update the event log by inserting $(1, \text{infection}, 0)$ and $(1, \text{onset}, d_1)$ into event history $\mathcal{A}$.
    
    \While{$\Psi$ is not empty and $t < T_f$}
        \State Pop the next infection attempt time $t_n$ from $\Psi$.
        \State Set $t = t_n$.
        \State Sample $u \sim \text{Uniform}(0,1)$.
        \If{$u < S(t)/(N-1)$}
            \State Select index $j$ uniformly at random from $\{i : Y_i(t) = 0\}$.
            \State Set $Y_j(t) = 1$.
            \State Run Algorithm~\ref{alg:individual_simulator} for an infection with index $j$ at time $t$, returning infection attempt times $\boldsymbol{s}_{j}$, reported symptom onset time $d_j$, and $\nu_j(t)$.
            \State Push $\boldsymbol{s}_{j}$ onto $\Psi$ and store $\nu_j(t)$ in $\mathcal{B}$.
            \State Update the event log by inserting $(j, \text{infection}, t)$ and $(j, \text{onset}, d_j)$ into event history $\mathcal{A}$.
        \Else
            \State Discard the event.
        \EndIf
    \EndWhile
    
    \Statex \hspace{-\algorithmicindent}\textbf{Return:} Event log $\mathcal{A}$, and log-10 viral load trajectories $\left\{ \nu_j(t) \right\}_{j=1}^{M}$, where $M$ is the total number of infected individuals.

\end{algorithmic}
\end{algorithm}

\clearpage
\section{Jacobian for GMM}\label{app:jacobian_derivation}

The target density is the joint density $p \left( R_0, \delta, \rho, \tau \right)$ and the fitted GMM approximates the density
\[
p \left(
f_2\left(\log R_0, \log \delta, \log \rho, \tau \right)
\right) = p \left( f_1(\log R_0), f_1( \log \delta), f_1(\log \rho), f_1(\tau) \right).
\]
Let $z_j = \log{\theta_j}$ for one of the non-negative scalar within-host parameter. The inverse
transformation is given by
\[
\theta_j = f^{-1}(z_j)
= \exp\!\left( \tilde{m}_{z_j} + \tilde{\sigma}_{z_j} z_j \right).
\]
Note that our transformation maps the standardised log parameters to the raw scale, and so the Jacobian relates to the forward mapping and is given by 
\[
\dod{f_1(\theta_j)}{\theta_j}
= \frac{1}{\tilde{\sigma}_{z_j}\theta_j}.
\]
For the simpler parameter $\tau$, the Jacobian is given by
\[
\dod{f_1(\tau)}{\tau} = \frac{1}{\tilde{\sigma}_{\tau}}.
\]
Since the transformation acts independently on each parameter, the Jacobian of the full transformation is diagonal and its determinant is given by
\[
\left| \det J_{f_2}(\boldsymbol{\theta}, 
\tau) \right|
= \frac{1}{\tilde{\sigma}_{\log R_0}\tilde{\sigma}_{\log\delta} \tilde{\sigma}_{\log\rho} \tilde{\sigma}_{\tau}R_0 \delta \rho}.
\]
The induced density on the raw parameters Eq.~\eqref{eq:within_host_surrogate} follows from the change of variables formula. 

\section{Between-host inference method details}\label{app:inference_method_details}

The target distribution is
\begin{equation}
\begin{aligned}
    & p \left( 
    \left\{ \Theta_{h}^{(B)}, \boldsymbol{u}_{h}^{(B)}, 
    \boldsymbol{\tau}_{h}^{(B)}, \boldsymbol{\omega}_{h} \right\}_{h \in \mathcal{H}},
    \boldsymbol{\zeta}
    \,\middle|\,
    \left\{ \boldsymbol{d}_{h} \right\}_{h \in \mathcal{H}}
    \right) \\ 
    &\propto \prod_{h \in \mathcal{H}} \Bigg[
    p \left( \boldsymbol{d}_{h} \,\middle|\,
    \Theta_{h}^{(B)}, \boldsymbol{u}_{h}^{(B)}, 
    \boldsymbol{\tau}_{h}^{(B)}, \boldsymbol{\omega}_{h},
    \boldsymbol{\zeta}
    \right) \\ 
    &\quad\times p \left(\boldsymbol{u}_{h}^{(B)}\,\middle|\,\Theta_{h}^{(B)}, \boldsymbol{\tau}_{h}^{(B)},\boldsymbol{\omega}_{h}, \boldsymbol{\zeta} \right) \\
    &\quad\times
    \hat{p}\!\left(\Theta_{h}^{(B)}, \boldsymbol{\tau}_{h}^{(B)}\right)
    p \left( \boldsymbol{\omega}_{h} \mid \boldsymbol{\zeta} \right)
    p(\boldsymbol{\zeta})\Bigg],
\end{aligned}
\end{equation}
where \(\hat{p}(\Theta_h^{(B)}, \boldsymbol{\tau}_h^{(B)})\) denotes the surrogate distribution learned from the first-stage within-host inference and is treated as fixed in the second stage.

Posterior sampling is performed using a Metropolis-within-Gibbs scheme that exploits the conditional independence structure induced by the household decomposition.
At each MCMC iteration, we cycle sequentially over households \(h \in \mathcal{H}\), updating household-specific latent variables and parameters conditional on the current value of the shared between-host parameters \(\boldsymbol{\zeta}\), followed by a global update of \(\boldsymbol{\zeta}\) conditional on all households.

Specifically, for each household \(h \in \mathcal{H}\), we sample household-level quantities from the conditional distribution
\begin{equation}
\begin{aligned}
    &p \left( 
    \Theta_{h}^{(B)}, \boldsymbol{u}_{h}^{(B)}, 
    \boldsymbol{\tau}_{h}^{(B)}, \boldsymbol{\omega}_{h}
    \,\middle|\,
    \boldsymbol{d}_{h}, \boldsymbol{\zeta}
    \right) \\ 
    &\propto
    p \left( \boldsymbol{d}_{h} \,\middle|\,
    \Theta_{h}^{(B)}, \boldsymbol{u}_{h}^{(B)}, 
    \boldsymbol{\tau}_{h}^{(B)}, \boldsymbol{\omega}_{h},
    \boldsymbol{\zeta}
    \right) \\ 
    &\quad\times p \left(
    \boldsymbol{u}_{h}^{(B)}
    \,\middle|\,
    \Theta_{h}^{(B)}, \boldsymbol{\tau}_{h}^{(B)},
    \boldsymbol{\omega}_{h}, \boldsymbol{\zeta}
    \right) \\ 
    &\quad\times \hat{p}\!\left(\Theta_{h}^{(B)}, \boldsymbol{\tau}_{h}^{(B)}\right)
    p \left( \boldsymbol{\omega}_{h} \mid \boldsymbol{\zeta} \right),
\end{aligned}
\end{equation}
where terms constant with respect to the household-specific parameters have been omitted.
Within each household block, latent infection times \(\boldsymbol{u}_{h}^{(B)}\) are updated jointly with other household-level quantities using Metropolis--Hastings proposals.

Conditional on the updated household-level quantities, we then sample the shared between-host parameters \(\boldsymbol{\zeta}\) from
\begin{equation}
\begin{aligned}
    & p \left( 
    \boldsymbol{\zeta}
    \,\middle|\,
    \left\{ \boldsymbol{d}_{h}, \Theta_{h}^{(B)}, \boldsymbol{u}_{h}^{(B)},
    \boldsymbol{\tau}_{h}^{(B)}, \boldsymbol{\omega}_{h} \right\}_{h \in \mathcal{H}}
    \right) \\ 
    &\propto
    \prod_{h \in \mathcal{H}} \Bigg[
    p \left( \boldsymbol{d}_{h} \,\middle|\,
    \Theta_{h}^{(B)}, \boldsymbol{u}_{h}^{(B)}, 
    \boldsymbol{\tau}_{h}^{(B)}, \boldsymbol{\omega}_{h},
    \boldsymbol{\zeta}
    \right) \\
    &\quad\times
    p \left(
    \boldsymbol{u}_{h}^{(B)}
    \,\middle|\,
    \Theta_{h}^{(B)}, \boldsymbol{\tau}_{h}^{(B)},
    \boldsymbol{\omega}_{h}, \boldsymbol{\zeta}
    \right)
    p(\boldsymbol{\zeta}) \Bigg],
\end{aligned}
\end{equation}
where all household-specific quantities are treated as fixed.

We use multivariate Gaussian block proposals for both household-level updates and the shared between-host parameters.
Proposal covariances are tuned using short pilot runs to achieve stable acceptance rates.

\section{Predictive distributions}\label{app:predictive_distributions} 

Suppose we wish to sample from the predictive distribution for $d^\star$. 
This could be the prior predictive, $p(d^\star)$, or the posterior predictive, $p(d^\star \mid \{\boldsymbol{d}_h, \mathcal{Y}_h\}_{h \in \mathcal{H}})$. 
For notational simplicity we will derive the prior predictive but note that the posterior predictive can be substituted in for the sampling distributions. 

The prior predictive density is given by 
\begin{equation*}
    p(d^\star) = \int p(d^\star \mid \boldsymbol{\theta}, \boldsymbol{\phi}, \tau, \boldsymbol{\zeta}, \omega) p(\boldsymbol{\theta}, \boldsymbol{\phi}, \tau, \boldsymbol{\zeta}, \omega) \dif \boldsymbol{\theta} \dif \boldsymbol{\phi} \dif \tau \dif \boldsymbol{\zeta} \dif \omega.
\end{equation*}
By construction of our model, we have 
\begin{equation*}
    \begin{aligned}
        p(d^\star) = \int & \bigg[ p(d^\star \mid \boldsymbol{\theta}, \boldsymbol{\phi}, \tau, \boldsymbol{\zeta}, \omega) \\
        & \times p(\tau \mid \boldsymbol{\theta}) p(\boldsymbol{\theta} \mid \boldsymbol{\phi}) p(\boldsymbol{\phi}) p(\omega \mid \boldsymbol{\zeta}) p(\boldsymbol{\zeta}) \bigg] \dif \boldsymbol{\theta} \dif \boldsymbol{\phi} \dif \tau \dif \boldsymbol{\zeta} \dif \omega.
    \end{aligned}
\end{equation*}
Monte-carlo estimates of this integral can be obtained by the following method:

\begin{enumerate}
    \item Draw $\boldsymbol{\phi} \sim p(\boldsymbol{\phi})$
    \item Draw $\boldsymbol{\theta} \mid \boldsymbol{\phi} \sim p(\boldsymbol{\theta} \mid \boldsymbol{\phi})$
    \item Draw $\tau \mid \boldsymbol{\theta} \sim p(\tau \mid \boldsymbol{\theta})$
    \item Draw $\boldsymbol{\zeta} \sim p(\boldsymbol{\zeta})$
    \item Draw $\omega \mid \boldsymbol{\zeta} \sim p(\omega \mid \boldsymbol{\zeta})$
    \item Solve the model to obtain $V(t)$, and construct $\lambda(t)$ as per Eq.~\eqref{eq:lambda}. 
\end{enumerate}

We construct our predictive distribution for the reported onset time by sampling this way and solving for $d^\star$ as described in Section~\ref{sec:data} of the main text. 
Noting that $d^\star$ is equivalent to solving the model with $u = 0$ (i.e. then $d^\star$ is the incubation period) as we did in Section~\ref{sec:pred_distributions}. 

\clearpage
\section{Viral kinetic trajectories for other individuals}\label{app:predictive_vl_distributions} 

\begin{figure}[!ht]
    \centering
    \includegraphics{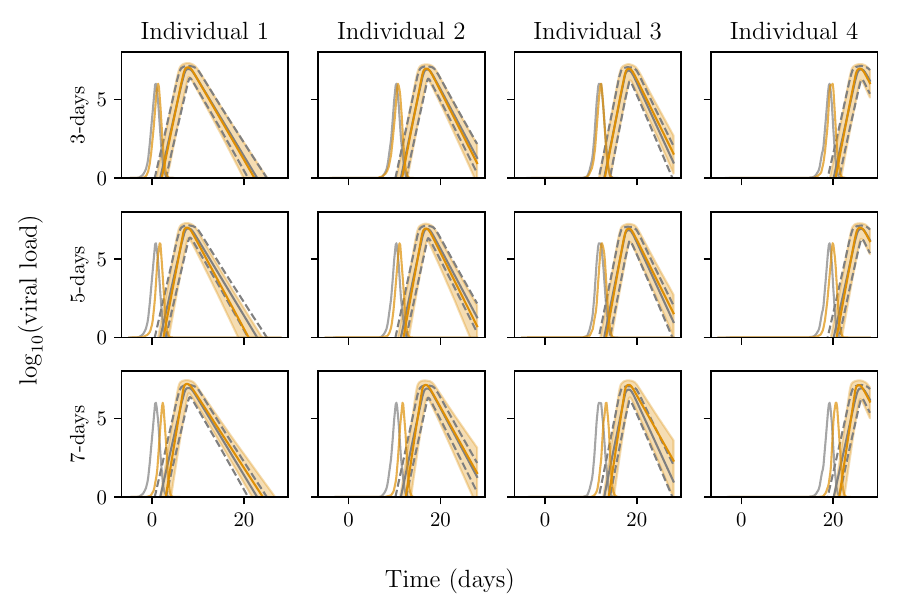}
    \caption{
        Posterior predictive viral load trajectories for individuals in the example dataset under the different sampling regimes. 
        Orange lines correspond to the specific sampling regime and grey lines correspond to the daily sampling as a reference. 
        The non-daily sampling regimes have the 95\% prediction intervals shown in orange and the daily sampling is not shaded, but is instead indicated with dashed lines. 
        The row indicates the sampling regime, indicated on the left-hand side.
        Individual indices are given in the column titles.
        The marginal posterior for the infection times are indicated by kernel density estimates near the start of the trajectories, colour coded to the relevant case. 
        Note that individual 1 is shown in the main text in Fig.~\ref{fig:household8_vl_predictions}.
    }
    \label{fig:household8_vl_predictions_app}
\end{figure}

\section{Computational performance}\label{app:performance}

We evaluate the computational performance of the model and its scaling behaviour as a function of dataset size, measured by the number of households $H$, using two metrics: wall-clock runtime and effective sample size (ESS) for a fixed number of iterations.

For each dataset size, we report four quantities: the runtime and ESS for both the within-host and between-host inference stages.
Reported runtimes correspond to a single dataset at each household size and include four executions of the MCMC subroutine as described in Section~\ref{sec:simulation_studies} (one pilot run and three sampling chains).
For the within-host model, chains are run sequentially as they are parallelised within the likelihood calculation. 
The between-host model is run in parallel across chains. 
Pilot runs for each component are a single chain and run prior to the inference itself. 
Runtimes were not averaged across datasets, as datasets with the same number of households may contain different numbers of individuals and therefore incur different computational costs.
In this example dataset there were 186 individuals.

Table~\ref{tbl:runtime_ess} shows the runtime and ESS results for an example dataset at each household size.
For the stage 1 model, runtime scales approximately linearly in $H$.
In stage 2, runtime also exhibits near-linear scaling but is longer overall because the sampling chains are twice as long by design.
In addition, the pilot chain in stage 2 is more computationally expensive, as the between-host sampler is run in serial and does not parallelise likelihood evaluations.
In both stages, the main computational bottleneck is solving the ODEs, making this linear scaling expected.
ESS values remain relatively stable across dataset sizes, indicating reasonable sampling efficiency as $H$ increases.

The overall computational cost of the method scales as $\mathcal{O}(\text{total number of individuals})$, and therefore approximately $\mathcal{O}(H)$ as household sizes are bounded.
The cost of approximating the posterior from the within-host model in order to pass forward to the between-host model is negligible (averaging only a few seconds per individual) and is omitted from reported runtimes; this computation can be parallelised easily as there is no dependence across individuals.
We also exclude runtimes for learning the timeshift distributions, as this is a one-time upfront cost that can be reused in both model stages as in \citet{morrisRandomTimeshiftApproximation2025a}.
Although this analysis does not constitute a comprehensive study of computational complexity, it illustrates the expected qualitative scaling behaviour of the method as the amount of data increases.
This demonstrates that the model is computationally feasible for practical implementation, even without additional resources such as those provided by high performance computing. 

\begin{table}[!ht]
    \centering
    \label{tbl:runtime_ess}
    \begin{tabular}{ccccc}
    \toprule
    $H$ &
    Runtime stage 1 (min) &
    Runtime stage 2 (min) &
    ESS stage 1 &
    ESS stage 2 \\
    \midrule
    $10$  & 5  & 12  & 1110 & 1515 \\
    $20$  & 8  & 21  & 1110 & 1470 \\
    $50$  & 17 & 54  & 753  & 1317 \\
    $100$ & 31 & 125 & 822  & 1671 \\
    \bottomrule
    \end{tabular}
    \caption{
        Computational performance by number of households ($H$) for within-host (stage 1) and between-host (stage 2) inference stages. 
        Runtimes are the total wall-clock time across four MCMC chains (one pilot run and three sampling chains), reported in minutes. 
        Effective sample sizes (ESS) report the minimum across all parameters within each stage, and are included to allow a broad comparison of relative sampling efficiency.
    }
\end{table}


\end{appendices}

\bibliographystyle{elsarticle-harv} 
\bibliography{sn-bibliography}

@book{gelmanBayesianDataAnalysis,
  title = {Bayesian Data Analysis},
  author = {Gelman, Andrew and Carlin, John B and Stern, Hal S and Dunson, David B and Vehtari, Aki and Rubin, Donald B},
  edition = {3rd},
  year = {2014},
  publisher = {CRC Press}
}

@article{pasupathy2011,
  title = {Generating {{Nonhomogeneous Poisson Processes}}},
  author = {Pasupathy, R.},
  year = 2011,
  journal = {Wiley Encyclopedia of Operations Research and Management Science},
  publisher = {Wiley-Blackwell},
  doi = {10.1002/9780470400531.eorms0356}
}

@article{allenPrimerStochasticEpidemic2017,
  title = {A Primer on Stochastic Epidemic Models: {{Formulation}}, Numerical Simulation, and Analysis},
  shorttitle = {A Primer on Stochastic Epidemic Models},
  author = {Allen, Linda J. S.},
  year = 2017,
  month = may,
  journal = {Infectious Disease Modelling},
  volume = {2},
  pages = {128--142},
  issn = {2468-0427},
  doi = {10.1016/j.idm.2017.03.001}
}

@article{swallowChallengesEstimationUncertainty2022,
  title = {Challenges in Estimation, Uncertainty Quantification and Elicitation for Pandemic Modelling},
  author = {Swallow, Ben and Birrell, Paul and Blake, Joshua and Burgman, Mark and Challenor, Peter and Coffeng, Luc E. and Dawid, Philip and De Angelis, Daniela and Goldstein, Michael and Hemming, Victoria and Marion, Glenn and McKinley, Trevelyan J. and Overton, Christopher E. and {Panovska-Griffiths}, Jasmina and Pellis, Lorenzo and Probert, Will and Shea, Katriona and Villela, Daniel and Vernon, Ian},
  year = 2022,
  month = mar,
  journal = {Epidemics},
  volume = {38},
  pages = {100547},
  issn = {1755-4365},
  doi = {10.1016/j.epidem.2022.100547},
  pmcid = {PMC7612598},
  pmid = {35180542}
}

@article{maEstimatingEpidemicExponential2020,
  title = {Estimating Epidemic Exponential Growth Rate and Basic Reproduction Number},
  author = {Ma, Junling},
  year = 2020,
  month = jan,
  journal = {Infectious Disease Modelling},
  volume = {5},
  pages = {129--141},
  issn = {2468-0427},
  doi = {10.1016/j.idm.2019.12.009}
}

@article{marionModellingUnderstandingPandemics2022,
  title = {Modelling: {{Understanding}} Pandemics and How to Control Them},
  shorttitle = {Modelling},
  author = {Marion, Glenn and Hadley, Liza and Isham, Valerie and Mollison, Denis and {Panovska-Griffiths}, Jasmina and Pellis, Lorenzo and Tomba, Gianpaolo Scalia and Scarabel, Francesca and Swallow, Ben and Trapman, Pieter and Villela, Daniel},
  year = 2022,
  month = jun,
  journal = {Epidemics},
  volume = {39},
  pages = {100588},
  issn = {1755-4365},
  doi = {10.1016/j.epidem.2022.100588}
}

@article{gariraPrimerMultiscaleModelling2018,
  title = {A Primer on Multiscale Modelling of Infectious Disease Systems},
  author = {Garira, Winston},
  year = 2018,
  month = sep,
  journal = {Infectious Disease Modelling},
  volume = {3},
  pages = {176--191},
  issn = {2468-2152},
  doi = {10.1016/j.idm.2018.09.005},
  pmcid = {PMC6326222},
  pmid = {30839905}
}

@article{perelsonModellingViralImmune2002,
  title = {Modelling Viral and Immune System Dynamics},
  author = {Perelson, Alan S.},
  year = 2002,
  month = jan,
  journal = {Nature Reviews Immunology},
  volume = {2},
  pages = {28--36},
  publisher = {Nature Publishing Group},
  issn = {1474-1741},
  doi = {10.1038/nri700},
  copyright = {2002 Springer Nature Limited}
}

@book{Cormen2022,
  title = {Introduction to Algorithms},
  author = {Cormen, Thomas H. and Leiserson, Charles Eric and Rivest, Ronald Linn and Stein, Clifford},
  year = 2022,
  edition = {Fourth edition},
  publisher = {The MIT Press},
  address = {Cambridge, Massachusetts London, England},
  isbn = {978-0-262-04630-5 978-0-262-36750-9},
  langid = {english}
}

@article{marcatoLearningsAustralianFirst2022,
  title = {Learnings from the {{Australian}} First Few {{X}} Household Transmission Project for {{COVID-19}}},
  author = {Marcato, Adrian J. and Black, Andrew J. and Walker, Camelia R. and Morris, Dylan and Meagher, Niamh and Price, David J. and McVernon, Jodie},
  year = 2022,
  month = nov,
  journal = {The Lancet Regional Health - Western Pacific},
  volume = {28},
  issn = {26666065},
  doi = {10.1016/j.lanwpc.2022.100573}
}

@misc{householdid,
  title = {Australian Household Size Profile},
  author = {{.id (informed decisions)}},
  howpublished = {\url{https://profile.id.com.au/australia/household-size}},
  note = {Accessed: 22 January 2026},
  year = {2022}
}

@article{gelmanWeakConvergenceOptimal1997,
  title = {Weak Convergence and Optimal Scaling of Random Walk {{Metropolis}} Algorithms},
  author = {Gelman, A. and Gilks, W. R. and Roberts, G. O.},
  year = 1997,
  month = feb,
  journal = {The Annals of Applied Probability},
  volume = {7},
  pages = {110--120},
  publisher = {Institute of Mathematical Statistics},
  issn = {1050-5164, 2168-8737},
  doi = {10.1214/aoap/1034625254}
}

@misc{FirstFewCases2021,
  title = {The First Few {{X}} Cases and Contacts ({{FFX}}) Investigation Protocol for Coronavirus Disease 2019 ({{COVID-19}}), Version 2.2},
  author = {{World Health Organization}},
  year = {2021},
  howpublished = {\url{https://www.who.int/publications/i/item/the-first-few-x-cases-and-contacts-(-ffx)-investigation-protocol-for-coronavirus-disease-2019-(-covid-19)-version-2.2}}
}

@article{bedardOptimalAcceptanceRates2008,
  title = {Optimal Acceptance Rates for {{Metropolis}} Algorithms: {{Moving}} beyond 0.234},
  author = {B{\'e}dard, Myl{\`e}ne},
  year = 2008,
  month = dec,
  journal = {Stochastic Processes and their Applications},
  volume = {118},
  pages = {2198--2222},
  publisher = {Elsevier BV},
  doi = {10.1016/j.spa.2007.12.005}
}

@article{tatsukawaAgentbasedNestedModel2023,
  title = {An Agent-Based Nested Model Integrating within-Host and between-Host Mechanisms to Predict an Epidemic},
  author = {Tatsukawa, Yuichi and Arefin, Md Rajib and Kuga, Kazuki and Tanimoto, Jun},
  year = 2023,
  month = dec,
  journal = {PLOS ONE},
  volume = {18},
  pages = {e0295954},
  publisher = {Public Library of Science},
  issn = {1932-6203},
  doi = {10.1371/journal.pone.0295954}
}

@article{marcQuantifyingRelationshipSARSCoV22021,
  title = {Quantifying the Relationship between {{SARS-CoV-2}} Viral Load and Infectiousness},
  author = {Marc, Aur{\'e}lien and Kerioui, Marion and Blanquart, Fran{\c c}ois and Bertrand, Julie and Mitj{\`a}, Oriol and {Corbacho-Monn{\'e}}, Marc and Marks, Michael and Guedj, Jeremie},
  editor = {Cobey, Sarah E and {Van der Meer}, Jos W},
  year = 2021,
  month = sep,
  journal = {eLife},
  volume = {10},
  pages = {e69302},
  publisher = {eLife Sciences Publications, Ltd},
  issn = {2050-084X},
  doi = {10.7554/eLife.69302}
}

@article{marcatoOngoingValueFirst2022,
  title = {The Ongoing Value of First Few {{X}} Studies for {{COVID-19}} in the {{Western Pacific Region}}},
  author = {Marcato, Adrian J and Fielding, James E and Crooks, Kristy and Massey, Peter D and Le, Linh-Vi and Bergeri, Isabel and McVernon, Jodie},
  year = 2022,
  month = mar,
  journal = {Western Pacific Surveillance and Response Journal : WPSAR},
  volume = {13},
  pages = {1--3},
  issn = {2094-7321},
  doi = {10.5365/wpsar.2022.13.1.873},
  pmcid = {PMC8989625},
  pmid = {35402064}
}

@article{vehtariRankNormalizationFoldingLocalization2021,
  title = {Rank-{{Normalization}}, {{Folding}}, and {{Localization}}: {{An Improved R\textasciicircum}} for {{Assessing Convergence}} of {{MCMC}} (with {{Discussion}})},
  shorttitle = {Rank-{{Normalization}}, {{Folding}}, and {{Localization}}},
  author = {Vehtari, Aki and Gelman, Andrew and Simpson, Daniel and Carpenter, Bob and B{\"u}rkner, Paul-Christian},
  year = 2021,
  month = jun,
  journal = {Bayesian Analysis},
  volume = {16},
  pages = {667--718},
  publisher = {International Society for Bayesian Analysis},
  issn = {1936-0975, 1931-6690},
  doi = {10.1214/20-BA1221}
}

@article{bezansonJuliaFreshApproach2017,
  title = {Julia: {{A Fresh Approach}} to {{Numerical Computing}}},
  shorttitle = {Julia},
  author = {Bezanson, Jeff and Edelman, Alan and Karpinski, Stefan and Shah, Viral B.},
  year = 2017,
  month = jan,
  journal = {SIAM Review},
  volume = {59},
  pages = {65--98},
  issn = {0036-1445, 1095-7200},
  doi = {10.1137/141000671}
}

@misc{leeuwenGaussianMixturesjl2026,
  title = {{{GaussianMixtures}}.Jl},
  author = {van Leeuwen, David},
  year = 2026,
  month = jan
}

@article{baccamKineticsInfluenzaVirus2006,
  title = {Kinetics of {{Influenza A Virus Infection}} in {{Humans}}},
  author = {Baccam, Prasith and Beauchemin, Catherine and Macken, Catherine A. and Hayden, Frederick G. and Perelson, Alan S.},
  year = 2006,
  month = aug,
  journal = {Journal of Virology},
  volume = {80},
  pages = {7590--7599},
  issn = {0022-538X, 1098-5514},
  doi = {10.1128/JVI.01623-05}
}

@article{morrisComputationRandomTimeshift2024,
  title = {Computation of Random Time-Shift Distributions for Stochastic Population Models},
  author = {Morris, Dylan and Maclean, John and Black, Andrew J.},
  year = 2024,
  month = aug,
  journal = {Journal of Mathematical Biology},
  volume = {89},
  pages = {33},
  issn = {1432-1416},
  doi = {10.1007/s00285-024-02132-6}
}

@article{liuGeneralFrameworkCutting2025,
  title = {A General Framework for Cutting Feedback within Modularized {{Bayesian}} Inference},
  author = {Liu, Yang and Goudie, Robert J B},
  year = 2025,
  month = sep,
  journal = {Journal of the Royal Statistical Society Series B: Statistical Methodology},
  volume = {87},
  pages = {1171--1199},
  issn = {1369-7412},
  doi = {10.1093/jrsssb/qkaf012}
}

@article{morrisRandomTimeshiftApproximation2025a,
  title = {Random Time-Shift Approximation Enables Hierarchical {{Bayesian}} Inference of Mechanistic within-Host Viral Dynamics Models on Large Datasets},
  author = {Morris, Dylan J. and Kennedy, Lauren and Black, Andrew J.},
  year = 2025,
  month = dec,
  journal = {PLOS Computational Biology},
  volume = {21},
  pages = {e1013775},
  publisher = {Public Library of Science},
  issn = {1553-7358},
  doi = {10.1371/journal.pcbi.1013775}
}

@incollection{reynoldsGaussianMixtureModels2009,
  title = {Gaussian {{Mixture Models}}},
  booktitle = {Encyclopedia of {{Biometrics}}},
  author = {Reynolds, Douglas},
  year = 2009,
  pages = {659--663},
  publisher = {Springer},
  address   = {Boston, MA},
  doi = {10.1007/978-0-387-73003-5_196},
  isbn = {978-0-387-73003-5}
}

@article{schultzeCOVID19HumanInnate2021,
  title = {{{COVID-19}} and the Human Innate Immune System},
  author = {Schultze, Joachim L. and Aschenbrenner, Anna C.},
  year = 2021,
  month = apr,
  journal = {Cell},
  volume = {184},
  pages = {1671--1692},
  issn = {0092-8674},
  doi = {10.1016/j.cell.2021.02.029},
  pmcid = {PMC7885626},
  pmid = {33743212}
}

@article{margiottaInvestigatingRelationshipImmune2025,
  title = {Investigating the Relationship between the Immune Response and the Severity of {{COVID-19}}: A Large-Cohort Retrospective Study},
  shorttitle = {Investigating the Relationship between the Immune Response and the Severity of {{COVID-19}}},
  author = {Margiotta, Riccardo Giuseppe and Sozio, Emanuela and Del Ben, Fabio and Beltrami, Antonio Paolo and Cesselli, Daniela and Comar, Marco and Devito, Alessandra and Fabris, Martina and Curcio, Francesco and Tascini, Carlo and Sanguinetti, Guido},
  year = 2025,
  month = jan,
  journal = {Frontiers in Immunology},
  volume = {15},
  publisher = {Frontiers},
  issn = {1664-3224},
  doi = {10.3389/fimmu.2024.1452638}
}

@article{clearyUsingViralLoad2021,
  title = {Using Viral Load and Epidemic Dynamics to Optimize Pooled Testing in Resource-Constrained Settings},
  author = {Cleary, Brian and Hay, James A. and Blumenstiel, Brendan and Harden, Maegan and Cipicchio, Michelle and Bezney, Jon and Simonton, Brooke and Hong, David and Senghore, Madikay and Sesay, Abdul K. and Gabriel, Stacey and Regev, Aviv and Mina, Michael J.},
  year = 2021,
  month = apr,
  journal = {Science Translational Medicine},
  volume = {13},
  pages = {eabf1568},
  issn = {1946-6234},
  doi = {10.1126/scitranslmed.abf1568},
  pmcid = {PMC8099195},
  pmid = {33619080}
}

@article{smithInfluenzaVirusInfection2018,
  title = {Influenza {{Virus Infection Model With Density Dependence Supports Biphasic Viral Decay}}},
  author = {Smith, Amanda P. and Moquin, David J. and Bernhauerova, Veronika and Smith, Amber M.},
  year = 2018,
  journal = {Frontiers in Microbiology},
  volume = {9},
  issn = {1664-302X}
}

@article{keVivoKineticsSARSCoV22021,
  title = {In Vivo Kinetics of {{SARS-CoV-2}} Infection and Its Relationship with a Person's Infectiousness},
  author = {Ke, Ruian and Zitzmann, Carolin and Ho, David D. and Ribeiro, Ruy M. and Perelson, Alan S.},
  year = 2021,
  month = dec,
  journal = {Proceedings of the National Academy of Sciences},
  volume = {118},
  pages = {e2111477118},
  issn = {0027-8424, 1091-6490},
  doi = {10.1073/pnas.2111477118}
}

@article{zitzmannHowRobustAre2024,
  title = {How Robust Are Estimates of Key Parameters in Standard Viral Dynamic Models?},
  author = {Zitzmann, Carolin and Ke, Ruian and Ribeiro, Ruy M. and Perelson, Alan S.},
  year = 2024,
  month = apr,
  journal = {PLOS Computational Biology},
  volume = {20},
  pages = {e1011437},
  publisher = {Public Library of Science},
  issn = {1553-7358},
  doi = {10.1371/journal.pcbi.1011437}
}

@article{barbourEscapeBoundaryMarkov2015,
  title = {Escape from the Boundary in {{Markov}} Population Processes},
  author = {Barbour, A. D. and Hamza, K. and Kaspi, Haya and Klebaner, F. C.},
  year = 2015,
  month = dec,
  journal = {Advances in Applied Probability},
  volume = {47},
  pages = {1190--1211},
  publisher = {Cambridge University Press},
  issn = {0001-8678, 1475-6064},
  doi = {10.1239/aap/1449859806}
}

@article{challengerModellingUpperRespiratory2022,
  title = {Modelling Upper Respiratory Viral Load Dynamics of {{SARS-CoV-2}}},
  author = {Challenger, Joseph D. and Foo, Cher Y. and Wu, Yue and Yan, Ada W. C. and Marjaneh, Mahdi Moradi and Liew, Felicity and Thwaites, Ryan S. and Okell, Lucy C. and Cunnington, Aubrey J.},
  year = 2022,
  month = dec,
  journal = {BMC Medicine},
  volume = {20},
  pages = {25},
  issn = {1741-7015},
  doi = {10.1186/s12916-021-02220-0}
}

@article{kisslerstephenm.ViralDynamicsSARSCoV22021,
  title = {Viral {{Dynamics}} of {{SARS-CoV-2 Variants}} in {{Vaccinated}} and {{Unvaccinated Persons}}},
  author = {Kissler, Stephen M. and Fauver, Joseph R. and Mack, Christina and Tai, Caroline G. and Breban, Mallery I. and Watkins, Anne E. and Samant, Radhika M. and Anderson, Deverick J. and Metti, Jessica and Khullar, Gaurav and Baits, Rachel and MacKay, Matthew and Salgado Daisy and Baker, Tim and Dudley, Joel T. and Mason, Christopher E. and Ho, David D. and Grubaugh, Nathan D. and Grad, Yonatan H.},
  year = 2021,
  month = dec,
  journal = {New England Journal of Medicine},
  volume = {385},
  pages = {2489--2491},
  publisher = {Massachusetts Medical Society},
  doi = {10.1056/NEJMc2102507}
}

@article{handelCrossingScaleWithinhost2015,
  title = {Crossing the Scale from Within-Host Infection Dynamics to between-Host Transmission Fitness: A Discussion of Current Assumptions and Knowledge},
  shorttitle = {Crossing the Scale from Within-Host Infection Dynamics to between-Host Transmission Fitness},
  author = {Handel, Andreas and Rohani, Pejman},
  year = 2015,
  month = aug,
  journal = {Philosophical Transactions of the Royal Society B: Biological Sciences},
  volume = {370},
  pages = {20140302},
  issn = {0962-8436},
  doi = {10.1098/rstb.2014.0302}
}

@article{kisslerViralDynamicsAcute2021,
  title = {Viral Dynamics of Acute {{SARS-CoV-2}} Infection and Applications to Diagnostic and Public Health Strategies},
  author = {Kissler, Stephen M. and Fauver, Joseph R. and Mack, Christina and Olesen, Scott W. and Tai, Caroline and Shiue, Kristin Y. and Kalinich, Chaney C. and Jednak, Sarah and Ott, Isabel M. and Vogels, Chantal B. F. and Wohlgemuth, Jay and Weisberger, James and DiFiori, John and Anderson, Deverick J. and Mancell, Jimmie and Ho, David D. and Grubaugh, Nathan D. and Grad, Yonatan H.},
  editor = {Riley, Steven},
  year = 2021,
  month = jul,
  journal = {PLOS Biology},
  volume = {19},
  pages = {e3001333},
  issn = {1545-7885},
  doi = {10.1371/journal.pbio.3001333}
}

@manual{stan2024,
  title = {Stan: A probabilistic programming language},
  author = {{Stan Development Team}},
  year = {2024},
  note = {Version 2.35},
  url = {https://mc-stan.org}
}

@misc{jax2018github,
  author = {James Bradbury and Roy Frostig and Peter Hawkins and Matthew James Johnson and Chris Leary and Dougal Maclaurin and George Necula and Adam Paszke and Jake Vander{P}las and Skye Wanderman-{M}ilne and Qiao Zhang},
  title = {{JAX}: composable transformations of {P}ython+{N}um{P}y programs},
  url = {http://github.com/jax-ml/jax},
  version = {0.3.13},
  year = {2018},
}

@article{svenssonNoteGenerationTimes2007,
  title = {A Note on Generation Times in Epidemic Models},
  author = {Svensson, Ake},
  year = 2007,
  month = jul,
  journal = {Mathematical Biosciences},
  volume = {208},
  pages = {300--311},
  issn = {0025-5564},
  doi = {10.1016/j.mbs.2006.10.010},
  pmid = {17174352}
}

@book{keelingModelingInfectiousDiseases2008,
  title = {Modeling {{Infectious}} {{Diseases}} in {{Humans}} and {{Animals}}},
  author = {Keeling, Matt J. and Rohani, Pejman},
  year = 2008,
  publisher = {Princeton University Press},
  address = {Princeton, NJ},
  isbn = {978-1-4008-4103-5}
}

@article{doranMathematicalMethodsScaling2023,
  title = {Mathematical Methods for Scaling from Within-Host to Population-Scale in Infectious Disease Systems},
  author = {Doran, James W. G. and Thompson, Robin N. and Yates, Christian A. and Bowness, Ruth},
  year = 2023,
  month = dec,
  journal = {Epidemics},
  volume = {45},
  pages = {100724},
  issn = {1755-4365},
  doi = {10.1016/j.epidem.2023.100724}
}

@article{puhachSARSCoV2ViralLoad2023,
  title = {{{SARS-CoV-2}} Viral Load and Shedding Kinetics},
  author = {Puhach, Olha and Meyer, Benjamin and Eckerle, Isabella},
  year = 2023,
  month = mar,
  journal = {Nature Reviews Microbiology},
  volume = {21},
  pages = {147--161},
  publisher = {Nature Publishing Group},
  issn = {1740-1534},
  doi = {10.1038/s41579-022-00822-w},
  copyright = {2022 Springer Nature Limited}
}

@article{almoceraMultiscaleModelWithinhost2018,
  title = {Multiscale Model Within-Host and between-Host for Viral Infectious Diseases},
  author = {Almocera, Alexis Erich S. and Nguyen, Van Kinh and {Hernandez-Vargas}, Esteban A.},
  year = 2018,
  month = oct,
  journal = {Journal of Mathematical Biology},
  volume = {77},
  pages = {1035--1057},
  issn = {1432-1416},
  doi = {10.1007/s00285-018-1241-y}
}

@article{goyalMultiscaleModellingReveals2022,
  title = {Multi-Scale Modelling Reveals That Early Super-Spreader Events Are a Likely Contributor to Novel Variant Predominance},
  author = {Goyal, Ashish and Reeves, Daniel B. and Schiffer, Joshua T.},
  year = 2022,
  month = apr,
  journal = {Journal of The Royal Society Interface},
  volume = {19},
  pages = {20210811},
  publisher = {Royal Society},
  doi = {10.1098/rsif.2021.0811}
}

@article{serenaReviewMultilevelModeling2023,
  title = {A Review of Multilevel Modeling and Simulation for Human Mobility and Behavior},
  author = {Serena, Luca and Marzolla, Moreno and D'Angelo, Gabriele and Ferretti, Stefano},
  year = 2023,
  month = sep,
  journal = {Simulation Modelling Practice and Theory},
  volume = {127},
  pages = {102780},
  issn = {1569-190X},
  doi = {10.1016/j.simpat.2023.102780}
}

@article{smithEfficientCouplingWithinand2025,
  title = {Efficient Coupling of Within-and between-Host Infectious Disease Dynamics},
  author = {Smith, Cameron A. and Ashby, Ben},
  year = 2025,
  month = apr,
  journal = {Journal of Theoretical Biology},
  volume = {602--603},
  pages = {112061},
  issn = {0022-5193},
  doi = {10.1016/j.jtbi.2025.112061}
}

@article{wangMultiscaleModelCOVID192022,
  title = {A {{Multiscale Model}} of {{COVID-19 Dynamics}}},
  author = {Wang, Xueying and Wang, Sunpeng and Wang, Jin and Rong, Libin},
  year = 2022,
  journal = {Bulletin of Mathematical Biology},
  volume = {84},
  pages = {99},
  issn = {0092-8240},
  doi = {10.1007/s11538-022-01058-8},
  pmcid = {PMC9360740},
  pmid = {35943625}
}

@article{houseInferringRisksCoronavirus2022,
  title = {Inferring Risks of Coronavirus Transmission from Community Household Data},
  author = {House, Thomas and Riley, Heather and Pellis, Lorenzo and Pouwels, Koen B and Bacon, Sebastian and Eidukas, Arturas and Jahanshahi, Kaveh and Eggo, Rosalind M and Sarah Walker, A.},
  year = 2022,
  month = sep,
  journal = {Statistical Methods in Medical Research},
  volume = {31},
  pages = {1738--1756},
  issn = {0962-2802, 1477-0334},
  doi = {10.1177/09622802211055853}
}

@article{blackCharacterisingPandemicSeverity2017,
  title = {Characterising Pandemic Severity and Transmissibility from Data Collected during First Few Hundred Studies},
  author = {Black, Andrew J. and Geard, Nicholas and McCaw, James M. and McVernon, Jodie and Ross, Joshua V.},
  year = 2017,
  month = jun,
  journal = {Epidemics},
  volume = {19},
  pages = {61--73},
  publisher = {Elsevier BV},
  doi = {10.1016/j.epidem.2017.01.004}
}

@article{mcleanPandemicH1N120092010,
  title = {Pandemic ({{H1N1}}) 2009 Influenza in the {{UK}}: Clinical and Epidemiological Findings from the First Few Hundred ({{FF100}}) Cases},
  shorttitle = {Pandemic ({{H1N1}}) 2009 Influenza in the {{UK}}},
  author = {McLean, E. and Pebody, R. G. and Campbell, C. and Chamberland, M. and Hawkins, C. and {Nguyen-Van-Tam}, J. S. and Oliver, I. and Smith, G. E. and Ihekweazu, C. and Bracebridge, S. and Maguire, H. and Harris, R. and Kafatos, G. and White, P. J. and {Wynne-Evans}, E. and Green, J. and Myers, R. and Underwood, A. and Dallman, T. and Wreghitt, T. and Zambon, M. and Ellis, J. and Phin, N. and Smyth, B. and McMenamin, J. and Watson, J. M.},
  year = 2010,
  month = nov,
  journal = {Epidemiology and Infection},
  volume = {138},
  pages = {1531--1541},
  issn = {0950-2688, 1469-4409},
  doi = {10.1017/S0950268810001366}
}

@article{boddingtonEpidemiologicalClinicalCharacteristics2021,
  title = {Epidemiological and Clinical Characteristics of Early {{COVID-19}} Cases, {{United Kingdom}} of {{Great Britain}} and {{Northern Ireland}}},
  author = {Boddington, Nicola L and Charlett, Andre and Elgohari, Suzanne and Byers, Chloe and Coughlan, Laura and Vilaplana, Tatiana Garcia and Whillock, Rosie and Sinnathamby, Mary and Panagiotopoulos, Nikolaos and Letley, Louise and MacDonald, Pauline and Vivancos, Roberto and Edeghere, Obaghe and Shingleton, Joseph and Bennett, Emma and Cottrell, Simon and McMenamin, Jim and Zambon, Maria and Ramsay, Mary and Dabrera, Gavin and Saliba, Vanessa and Bernal, Jamie Lopez},
  year = 2021,
  month = mar,
  journal = {Bulletin of the World Health Organization},
  volume = {99},
  pages = {178--189},
  issn = {0042-9686},
  doi = {10.2471/BLT.20.265603}
}

@article{ashcroftTesttraceisolatequarantineTTIQIntervention2022,
  title = {Test-Trace-Isolate-Quarantine ({{TTIQ}}) Intervention Strategies after Symptomatic {{COVID-19}} Case Identification},
  author = {Ashcroft, Peter and Lehtinen, Sonja and Bonhoeffer, Sebastian},
  editor = {Blumberg, Seth},
  year = 2022,
  month = feb,
  journal = {PLOS ONE},
  volume = {17},
  pages = {e0263597},
  issn = {1932-6203},
  doi = {10.1371/journal.pone.0263597}
}

@article{hayEstimatingEpidemiologicDynamics2021,
  title = {Estimating Epidemiologic Dynamics from Cross-Sectional Viral Load Distributions},
  author = {Hay, James A. and {Kennedy-Shaffer}, Lee and Kanjilal, Sanjat and Lennon, Niall J. and Gabriel, Stacey B. and Lipsitch, Marc and Mina, Michael J.},
  year = 2021,
  month = jul,
  journal = {Science},
  volume = {373},
  issn = {0036-8075, 1095-9203},
  doi = {10.1126/science.abh0635}
}

@article{martinez-corralHillFunctionUniversal2024,
  title = {The {{Hill}} Function Is the Universal {{Hopfield}} Barrier for Sharpness of Input--Output Responses},
  author = {{Martinez-Corral}, Rosa and Nam, Kee-Myoung and DePace, Angela H. and Gunawardena, Jeremy},
  year = 2024,
  month = may,
  journal = {Proceedings of the National Academy of Sciences},
  volume = {121},
  pages = {e2318329121},
  publisher = {Proceedings of the National Academy of Sciences},
  doi = {10.1073/pnas.2318329121}
}

@article{madewellHouseholdSecondaryAttack2022,
  title = {Household {{Secondary Attack Rates}} of {{SARS-CoV-2}} by {{Variant}} and {{Vaccination Status}}},
  author = {Madewell, Zachary J. and Yang, Yang and Longini, Ira M. and Halloran, M. Elizabeth and Dean, Natalie E.},
  year = 2022,
  month = apr,
  journal = {JAMA Network Open},
  volume = {5},
  pages = {e229317},
  issn = {2574-3805},
  doi = {10.1001/jamanetworkopen.2022.9317},
  pmcid = {PMC9051991},
  pmid = {35482308}
}

@article{heTemporalDynamicsViral2020,
  title = {Temporal Dynamics in Viral Shedding and Transmissibility of {{COVID-19}}},
  author = {He, Xi and Lau, Eric H. Y. and Wu, Peng and Deng, Xilong and Wang, Jian and Hao, Xinxin and Lau, Yiu Chung and Wong, Jessica Y. and Guan, Yujuan and Tan, Xinghua and Mo, Xiaoneng and Chen, Yanqing and Liao, Baolin and Chen, Weilie and Hu, Fengyu and Zhang, Qing and Zhong, Mingqiu and Wu, Yanrong and Zhao, Lingzhai and Zhang, Fuchun and Cowling, Benjamin J. and Li, Fang and Leung, Gabriel M.},
  year = 2020,
  month = may,
  journal = {Nature Medicine},
  volume = {26},
  pages = {672--675},
  publisher = {Nature Publishing Group},
  issn = {1546-170X},
  doi = {10.1038/s41591-020-0869-5},
  copyright = {2020 The Author(s), under exclusive licence to Springer Nature America, Inc.}
}

@article{yinAccurateStochasticSimulation2025,
  title = {Accurate Stochastic Simulation Algorithm for Multiscale Models of Infectious Diseases},
  author = {Yin, Yuan and Flegg, Jennifer A. and Flegg, Mark B.},
  year = 2025,
  month = sep,
  journal = {Journal of Theoretical Biology},
  volume = {612},
  pages = {112194},
  issn = {0022-5193},
  doi = {10.1016/j.jtbi.2025.112194}
}

@article{amosTutorialAmortizedOptimization2023,
  title = {Tutorial on {{Amortized Optimization}}},
  author = {Amos, Brandon},
  year = 2023,
  journal = {Foundations and Trends in Machine Learning},
  volume = {16},
  pages = {592--732},
  issn = {1935-8237, 1935-8245},
  doi = {10.1561/2200000102}
}

@incollection{marinoGeneralMethodAmortizing2018,
    title = {A {General Method} for {Amortizing Variational Filtering}},
    author = {Marino, Joseph and Cvitkovic, Milan and Yue, Yisong},
    booktitle = {Advances in {Neural Information Processing Systems}},
    year = {2018},
    volume = {31},
    publisher = {Curran Associates, Inc.},
    address = {Red Hook, NY, USA}
}

\end{document}